\documentclass[
final
]{dmtcs-episciences}

\usepackage[utf8]{inputenc}
\usepackage{amsmath,amssymb,amsfonts}
\usepackage{subcaption}
\usepackage{tikz}
\usepackage{enumerate}
\usepackage{comment}
\usepackage[thmmarks,amsmath]{ntheorem}

\usepackage{url}
\usepackage{hyperref}

\newcommand{\Z}{\mathbb{Z}}
\newcommand{\Q}{\mathbb{Q}}
\newcommand{\R}{\mathbb{R}}

\newcommand{\CC}{\mathcal{C}}
\newcommand{\F}{\mathbb{F}}
\newcommand{\PP}{\mathcal{P}}
\newcommand{\T}{\mathcal{T}}
\newcommand{\dist}{\text{dist}}

{\newtheorem{definition}{Definition}}
{\newtheorem{lemma}{Lemma}}
{\newtheorem{proposition}{Proposition}}
{\newtheorem{theorem}{Theorem}}
{\newtheorem{corollary}{Corollary}}

\usepackage[round]{natbib}
\title{Ammann Bars for Octagonal Tilings}
\author{Carole Porrier\affiliationmark{1}
  \and {Th}omas Fernique\affiliationmark{2}}
\affiliation{
  LIPN, Université Sorbonne Paris Nord, Villetaneuse, France\\
  Higher School of Economics, Moscow, Russia} 
\begin{document}

\publicationdata{vol. 26:3}{2024}{15}{10.46298/dmtcs.10764}{2023-01-03; 2023-01-03; 2024-07-26; 2024-09-26}{2024-10-04}
\maketitle

\vspace{-0.05cm}
\begin{abstract}
Ammann bars are formed by segments (decorations) on the tiles of a tiling such that forming straight lines with them while tiling forces non-periodicity.
Only a few cases are known, starting with Robert Ammann's observations on Penrose tilings, but there is no general explanation or construction. 
In this article we propose a general method for cut and project tilings based on \emph{subperiods} and we illustrate it with an aperiodic set of 36 decorated prototiles related to what we called \emph{Cyrenaic tilings}.
\end{abstract}


\section{Introduction}
\label{sec:intro}

\begin{figure}[ht]
    \centering
    \begin{tikzpicture}[scale=1.85]
    \path (0,0) edge (0.588,0.809) edge (0.588,-0.809);
    \path (1.176,0) edge (0.588,0.809) edge (0.588,-0.809);
    \path[dashed] (0.588,0.809) edge (0.588,-0.809);
    \draw (0.588,0.809) node[left] {$\frac{1}{4}$} ;
    \draw (0.147,-0.202) node[left] {$\frac{1}{2\varphi}$} ;
    \draw (1.1,0.1) node[above] {$\frac{\varphi}{2}$} ;
    \path[thick,color=orange] (0.588,0.618) edge (0.497,0.684) edge (0.294,0.4045) edge (0.679,0.684) edge (0.882,0.4045);
    \path[thick,color=orange] (0.147,-0.202) edge (0.588,-0.202) edge (0.294,0.4045);
    \path[thick,color=orange] (1.029,-0.202) edge (0.588,-0.202) edge (0.882,0.4045);
    \path[thick,color=blue] (0.658,-0.132) edge (0.588,-0.132) edge (0.658,-0.202);
    \path[thick,color=blue] (0.664,0.589) edge (0.622,0.646) edge (0.719,0.624); 
    
    \fill[color=blue!20] (-0.363,-1.4) -- (0.588,-1.709) -- (1.539,-1.4) -- (0.588,-1.091) -- cycle;
    \path (-0.363,-1.4) edge (0.588,-1.709) edge (0.588,-1.091);
    \path (1.539,-1.4) edge (0.588,-1.709) edge (0.588,-1.091);
    \path[dashed] (0.588,-1.709) edge (0.588,-1.091);
    \draw (0.35,-1.168) node[above] {$\frac{\varphi}{2}$} ;
    \draw (1.487,-1.4) node[above] {$\frac{1}{4}$} ;
    \path[thick,color=orange] (-0.1814,-1.459) edge (-0.216,-1.3523) edge (0.1125,-1.2455);
    \path[thick,color=orange] (1.3574,-1.459) edge (1.392,-1.3523) edge (1.0635,-1.2455);
    \path[thick,color=orange] (0.1125,-1.2455) edge (1.0635,-1.2455);
    \path[thick,color=blue] (0.658,-1.3155) edge (0.588,-1.3155) edge (0.658,-1.2455);
    \path[thick,color=blue] (-0.126,-1.4) edge (-0.146,-1.3323) edge (-0.196,-1.42); 
    
    \draw (0,-1.85) node {} ;
    \end{tikzpicture}
    \qquad
    \includegraphics[width=0.55\textwidth]{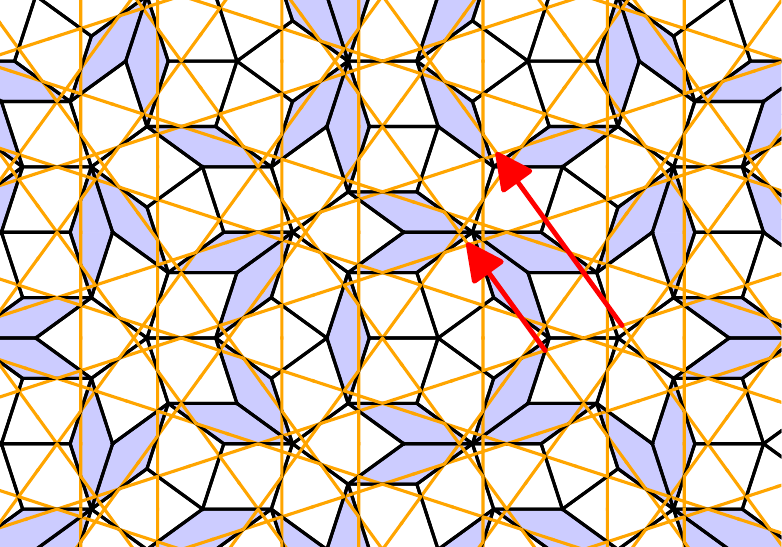}
    \caption{Left: Penrose tiles with Ammann segments (in orange).
    On each rhombus the dashed line is an axis of symmetry and the sides have length $\varphi=(1+\sqrt{5})/2$.
    Right: Ammann bars on a valid pattern of Penrose tiles, where each segment is correctly prolonged on adjacent tiles. The red vectors are ``integer versions'' of one \textit{subperiod}.}
    \label{segments}
\end{figure}
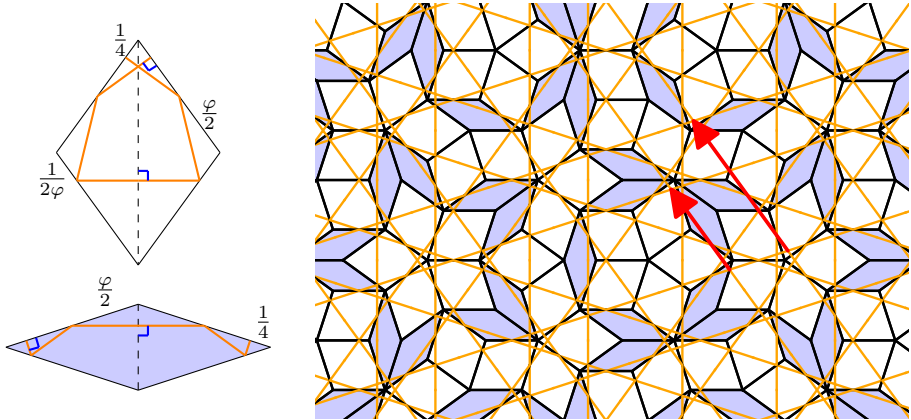

Shortly after the famous Penrose tilings were introduced by Roger \cite{penrose1974,penrose1978,penrose1979} and popularized by Martin \cite{gardner1977}, amateur mathematician Robert Ammann \citep{Senechal2004TheMM} found particularly interesting decorations of the tiles (Figure \ref{segments}): if one draws segments in the same way on all congruent tiles\footnote{taking into account the original markings on the sides that define the assembly rules.} then on any valid tiling all those segments compose continuous straight lines, going in five different directions. 
Conversely if one follows the assembly rule consisting of prolonging every segment on the tiles into a straight line then the obtained tiling is indeed a Penrose tiling.
Those lines are called \textit{Ammann bars} and the corresponding matching rule is locally equivalent to the ones given by Penrose using arrows on the sides or alternative decorations.\\

Penrose tilings have many interesting properties and can be generated in several ways.
The \textit{cut and project} method\footnote{Terms in italics which are not defined in the introduction are defined formally in further sections. The introduction is meant to give a general idea of the article.} follows their algebraic study by \cite{bruijn1981}.
\cite{Bee82} soon proposed a whole family of tilings based on it , including the Ammann-Beenker tilings that Ammann found independently.
A cut and project tiling can be seen as a discretization of a two-dimensional plane in an $n$-dimensional Euclidean space ($n>2$), and we will talk about \textit{$n\to2$ tilings} in that sense.
When the \textit{slope} of the plane does not contain any rational line, the tiling is non-periodic.
This is the case for Penrose tilings for instance, so the set of tiles defining them is \textit{aperiodic}: one can tile the plane with its tiles but only non-periodically.
The first aperiodic tileset was found by Berger, thus proving the undecidability of the \textsl{Domino Problem} \citep{Berger1966} and relating tilings to logic.
Since then, relatively few others were exhibited: many non-periodic tilings exist (even infinitely many using the cut and project method), but we usually do not have a corresponding aperiodic tileset.\\

Links were made between such tilings and quasicrystals \citep{Senechal1995,steinhardt1987}, that is crystals whose diffraction pattern is not periodic but still ordered, with rotational symmetries.
The study of \textit{local rules}, i.e. constraints on the way tiles can fit together in finite patterns, can help modeling the long range aperiodic order of quasicrystals.
For instance, Penrose tilings are defined by their 1-\textit{atlas} \citep{fernique-Lutfalla2023}, which is a small number of small patterns: any and all tilings containing only those patterns (of the given size) are Penrose tilings.
Alternately, they can also be defined by their \textit{Ammann local rules}, as stated in the first paragraph.
On the contrary, it was proven \citep{BF2015-4p} that Ammann-Beenker tilings, also known as 8-fold tilings, do not have \textit{weak local rules} (or ``colored'' local rules), i.e. no finite set of patterns (\textit{r-atlas}) is enough to characterize them.
\cite{socolar1989} found a variant of Ammann bars for them, but they extend outside the boundary of the tiles, thus do not fit the framework considered here.\\

In addition to finding a suitable tileset, one of the difficulties lies in proving that this tileset, with only nearest-neighbor rules, can enforce projection tilings with arbitrary slopes. 
\cite{GS1998-matchingRules} and then \cite{fernique-Ollinger2010} designed constructions in the case of substitutive tilings: they showed that, under specific conditions, substitutive tilings have colored nearest-neighbor matching rules. In other words, substitutions can be forced by colored local rules.
In the case of cut and project tilings, this is a corollary of a construction presented later by \cite{Fernique-Sablik2012} and based on works by \cite{Aubrun_2013}, and \cite{Hochman2009}.
This more general construction yields weak local rules for every computable slope, yet it is quite heavy, since it simulates Turing machines.
The main result presented here is actually a corollary of Fernique and Sablik’s, but the construction is completely different and much lighter because it is specific to a narrower class of tilings, for (\textit{a priori}) non-substitutive tilings, and additionally we exhibit more information on the structure of these tilings in the process.
Specifically, we provide sufficient conditions for some cut and project tilings to have Ammann bars.\\

\cite{grunbaum1987} detail the properties of Ammann bars in the case of Penrose tilings and their close relation to the \textsl{Fibonacci word},
which inspired Stehling to find an equivalent in 3-dimensional space \citep{Stehling1992}.
They also present two tilesets by Ammann with Ammann bars (A2 and A3) but these are substitutive and not cut and project tilings.
Generally speaking, many questions remain around Ammann bars and for now each family of aperiodic tilings has to be treated on a case-by-case basis,
as \cite{luck93} did with the examples of Ammann bars introduced by \cite{socolar1989} and \cite{INGALLS1993}.
Yet they can prove quite useful in the study of the structure of tilings, and were used by \cite{PB2020} to solve a combinatorial optimization problem on graphs defined by Penrose tilings.\\

When it comes to $4\to2$ tilings (discretizations\footnote{A discretization of a 2-plane $P$ is, roughly speaking, a set of unit facets of $\Z^4$ which approximate the continuous plane $P$. The formal notion is \textit{planar}, as defined in Section 2.} of 2-dimensional planes in $\R^4$) and a few others like Penrose, which are $5\to2$ tilings, the existence of weak local rules can be expressed in terms of \textit{subperiods}, which are particular vectors of the slope \citep{BF15,BF2017}. 
Careful examination of Penrose tilings from this angle shows that Ammann bars have the same directions as subperiods: 
there are two subperiods in each direction, one being $\varphi$ times longer than the other.
Additionally, the lengths of the ``integer versions'' of subperiods are closely related to the distances between two consecutive Ammann bars in a given direction, as can be seen in Figure \ref{segments} (more details are given in Appendix \ref{annexe-penrose}).
Though interesting, this special case is too particular to hope for a generalization from it alone.
Nonetheless, we think that Ammann bars are related to subperiods.\\

Since subperiods are simpler for $4\to2$ tilings, for which we also have a stronger result regarding weak local rules, we focus on those.
Namely, \cite{BF2017} showed that a $4\to2$ tiling has weak local rules if and only if its slope is characterized by its subperiods.
As mentioned above, Ammann-Beenker tilings have no local rules and their slope cannot be characterized by its subperiods.
The same holds for 12-fold tilings \citep{BF2015-4p} for which Socolar also designed Ammann lines \citep{socolar1989}.
This means that these two types of tilings have \textit{decorated} local rules, even though they have no weak local rules.
In both cases, the ``primary'' Ammann bars found by Socolar have the same directions as projected subperiods but tiles decorated with those lines alone allow for periodic tilings of the plane, which is why Socolar introduced ``secondary'' Ammann bars.
We conjecture that these lines, which are not related to subperiods, might force the correct frequencies of tiles, differentiating Ammann-Beenker tilings from other Beenker tilings, but we have not investigated further yet.\\

In any case, it seems some conditions of alignment play a part in the existence of Ammann bars.
This led us to introduce the notion of \textit{fine projection} (Def. \ref{def1} p.~\pageref{def1}) on a slope. 
We propose a constructive method -- the \textit{FP-method} (fine projection method) -- to find Ammann bars for $4\to2$ tilings which are characterized by subperiods and for which we can find a fine projection. 
The FP-method outputs a finite set of tiles, which we call an \textit{FP-tileset}.
We prove the following result:

\begin{theorem}\label{thm}
Any tileset obtained using the FP-method from a totally irrational slope is aperiodic.
\end{theorem}

\begin{figure}
    \centering
    \includegraphics[width=0.85\textwidth]{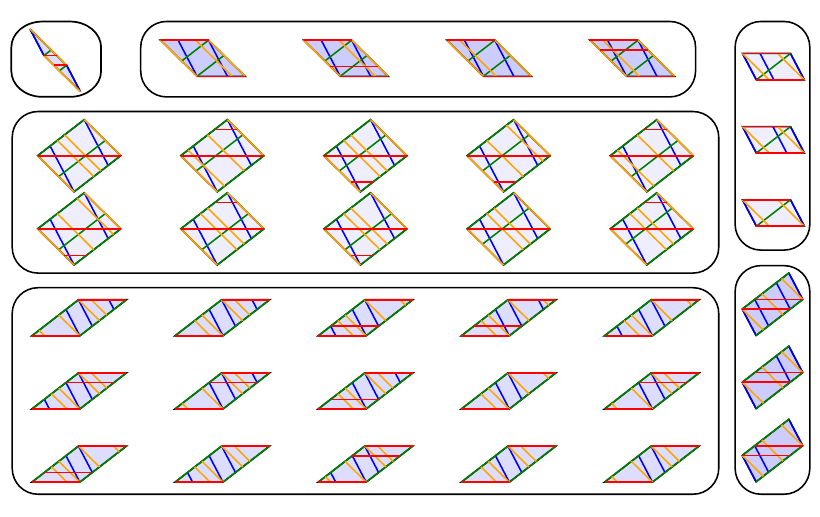}
    \caption{Set $\CC$ of 36 decorated prototiles obtained from Cyrenaic tilings. Any tiling by these tiles where segments extend to lines is non-periodic.}
    \label{decotiles}
\end{figure}

We initially searched for examples of slopes based on two criteria: the slope has to be quadratic to meet the algebraic constraints (being characterized by subperiods), and we wanted ``short'' subperiods to enable human-eye observation of the corresponding drawings.
The method involves a trial-and-error component: for each slope, we have to test whether it is possible to find a fine projection or not.
Our first try was on \textit{Golden Octagonal tilings} but their slope admits no fine projection.
Hence we randomly generated potential candidates, found a few having a fine projection, and proved the above result on a particular example which we presented in the LATIN 2022 conference \citep{pf2022}.
It consists of $4\to 2$ tilings defined by a given slope based on the irrationality of $\sqrt{3}$, that we called \textit{Cyrenaic tilings} in reference to Theodorus of Cyrene who proved $\sqrt{3}$ to be irrational.
In this case, the FP-method yields the set of decorated tiles depicted in Figure \ref{decotiles}.
Those tiles give Ammann bars to Cyrenaic tilings.
We use this example here to illustrate the general case.
We then found many examples of $4\to 2$ tilings characterized by their subperiods and admitting a fine projection, and we provide a program to generate the corresponding decorated tilesets for any slope meeting these conditions.
Note that there might be several fine projections for a given slope, hence different possible decorated tilesets depending on the choice we make at this step.\\

The case of Penrose indicates that our construction could (and should) be adapted in order to work for $5\rightarrow2$ tilings, or general cut and project ($n\to d$) tilings.
In particular, for Penrose the lines are shifted and the number of lines is reduced compared with our method, so that only two decorated tiles are needed.
Besides, in each direction the distance between two consecutive lines can take only two values, and the sequence of intervals is Sturmian and substitutive.
In the case of Cyrenaic tilings, it is an open question whether the bi-infinite word defined by each sequence of intervals between Ammann bars is substitutive. If so, it may be that they could be composed according to the substitution.
As for $n\to3$ tilings, the case of icosahedral Ammann rhombohedral tilings could be studied as a first example of ``Ammann planes'' (instead of lines), since their slope is characterized by subperiods \citep{BF2016icosahedral}.\\

More generally, we conjecture that the bi-infinite words defined by Ammann bars are either binary or ternary words, probably billiard words, in relation with the Three-Distance Theorem.
Such questions are still open, although there are clues as to how such properties could be explained.
Lines could also be shifted as in the case of Penrose tilings, instead of passing through vertices.
An optimal shift (reducing the number of lines or tiles) would then have to be determined.
Besides, it would be interesting to know whether the hypothesis of having a fine projection is necessary to have Ammann bars.
A link to our SageMath code as well as some more technical explanations are given in Appendix \ref{annexe-code}.\\

The paper is organized as follows. Section 2 introduces the setting, providing the necessary formal definitions, in particular local rules and subperiods. 
In Section 3 we present the FP-method to construct a set of decorated prototiles yielding Ammann bars. We rely on subperiods characterizing a slope as well as a fine projection, to prove that the obtained set is always finite.
Finally, in Section 4 we show that the Ammann segments on the tiles of an FP-tileset force the corresponding tilings to all have the same subperiods, thus proving Theorem \ref{thm}.

\section{setting} 
\label{sec:setting}

\subsection{Canonical cut and project tilings}
\label{sub:canonical}

A \textbf{tiling} of the plane is a covering by \textbf{tiles}, i.e. compact subsets of the space, whose interiors are pairwise disjoint. 
In this article we focus on \textbf{tilings by parallelograms}: 
let $v_0,...,v_{n-1}$ ($n\geq3$) be pairwise non-collinear vectors of the Euclidean plane, they define $n\choose 2$ parallelogram \textit{prototiles} which are the sets $T_{ij}:=\{\lambda v_i+\mu v_j\mid 0\leq\lambda,\mu\leq1\}$; 
then the tiles of a tiling by parallelograms are translated prototiles (tile rotation or reflection is forbidden), satisfying the edge-to-edge condition: the intersection of two tiles is either empty, a vertex or an entire edge.
When the $v_i$'s all have the same length, such tilings are called \textit{rhombus tilings}.

\begin{figure}[t]
    \centering
    \includegraphics[width=0.31\textwidth]{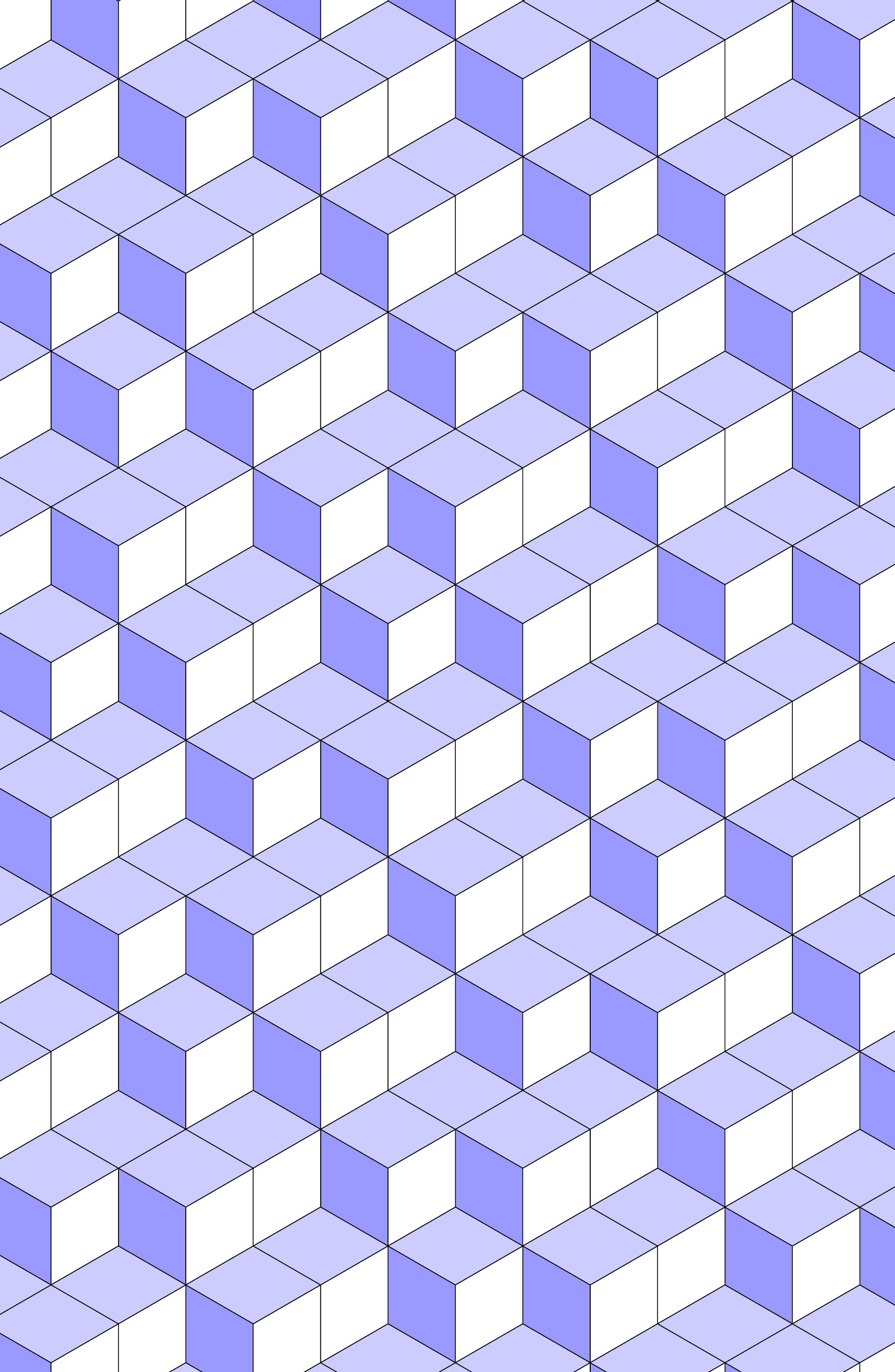}
    \hfill
    \includegraphics[width=0.31\textwidth]{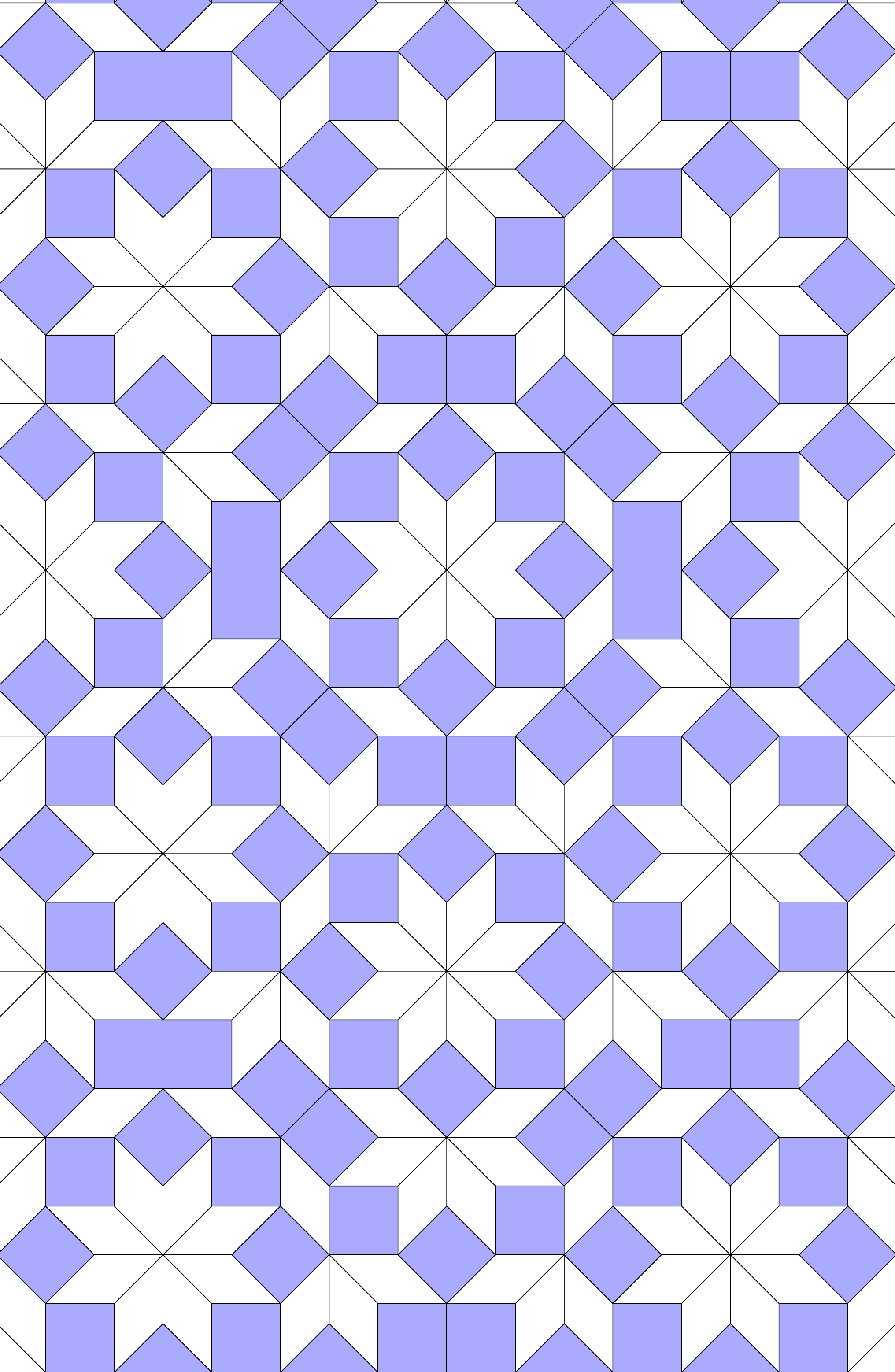}
    \hfill
    \includegraphics[width=0.31\textwidth]{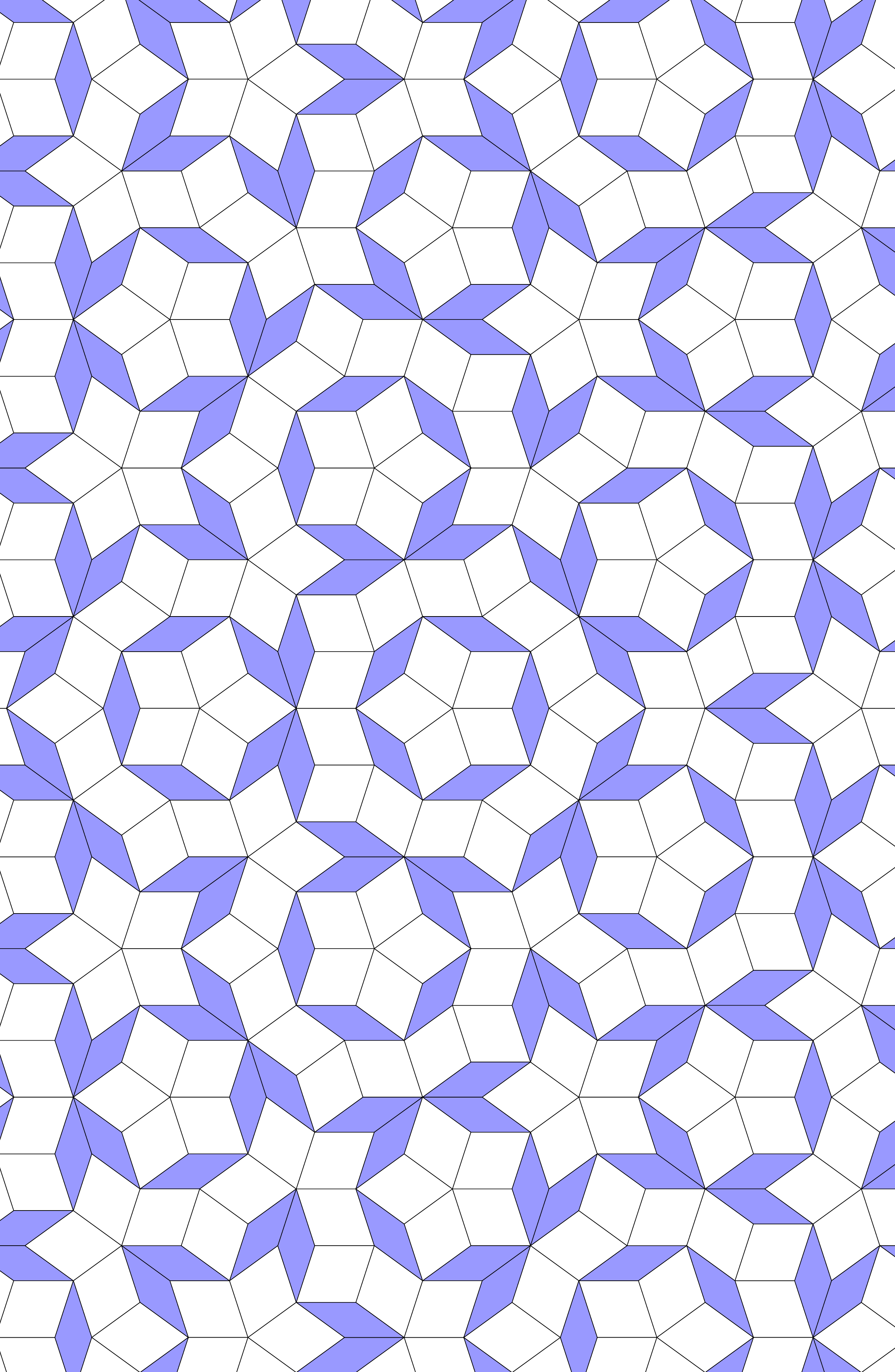}
    \caption{Examples. Left: Rauzy tiling from which you can visualize the lift in $\R^3$. Center: Ammann-Beenker tiling. Right: Penrose tiling.}
    \label{ex1}
\end{figure}

Let $e_0,...,e_{n-1}$ be the canonical basis of $\R^n$.
Following \cite{Levitov1988} and \cite{BF15},
a tiling by parallelograms can be \textbf{lifted} in $\R^n$, to correspond to a ``stepped'' surface of dimension 2 in $\R^n$,  which is unique up to the choice of an initial vertex. 
An arbitrary vertex is first mapped onto the origin, then each tile of type $T_{ij}$ is mapped onto the 2-dimensional face of a unit hypercube of $\Z^n$ generated by $e_i$ and $e_j$, such that two tiles adjacent along an edge $v_i$ are mapped onto two faces adjacent along an edge $e_i$.
This is particularly intuitive for $3\to 2$ tilings which are naturally seen in $3$ dimensions (Fig.~\ref{ex1}, left).
The principle is the same for larger $n$, though difficult to visualize.

If a tiling by parallelograms can be lifted into a tube $E+[0,t]^n$ where $E\subset\R^n$ is a plane and $t\geq1$, then this tiling is said to be \textbf{planar}.
In that case, \textbf{thickness} of the tiling is the smallest suitable $t$, and the corresponding $E$ is called the \textbf{slope} of the tiling (unique up to translation).
A planar tiling by parallelograms can thus be seen as an approximation of its slope, which is as good as the thickness is small.
Planarity is said \textbf{strong} if $t=1$ and \textbf{weak} otherwise.\\

\begin{figure}[ht]
    \centering
    \includegraphics[height=7cm]{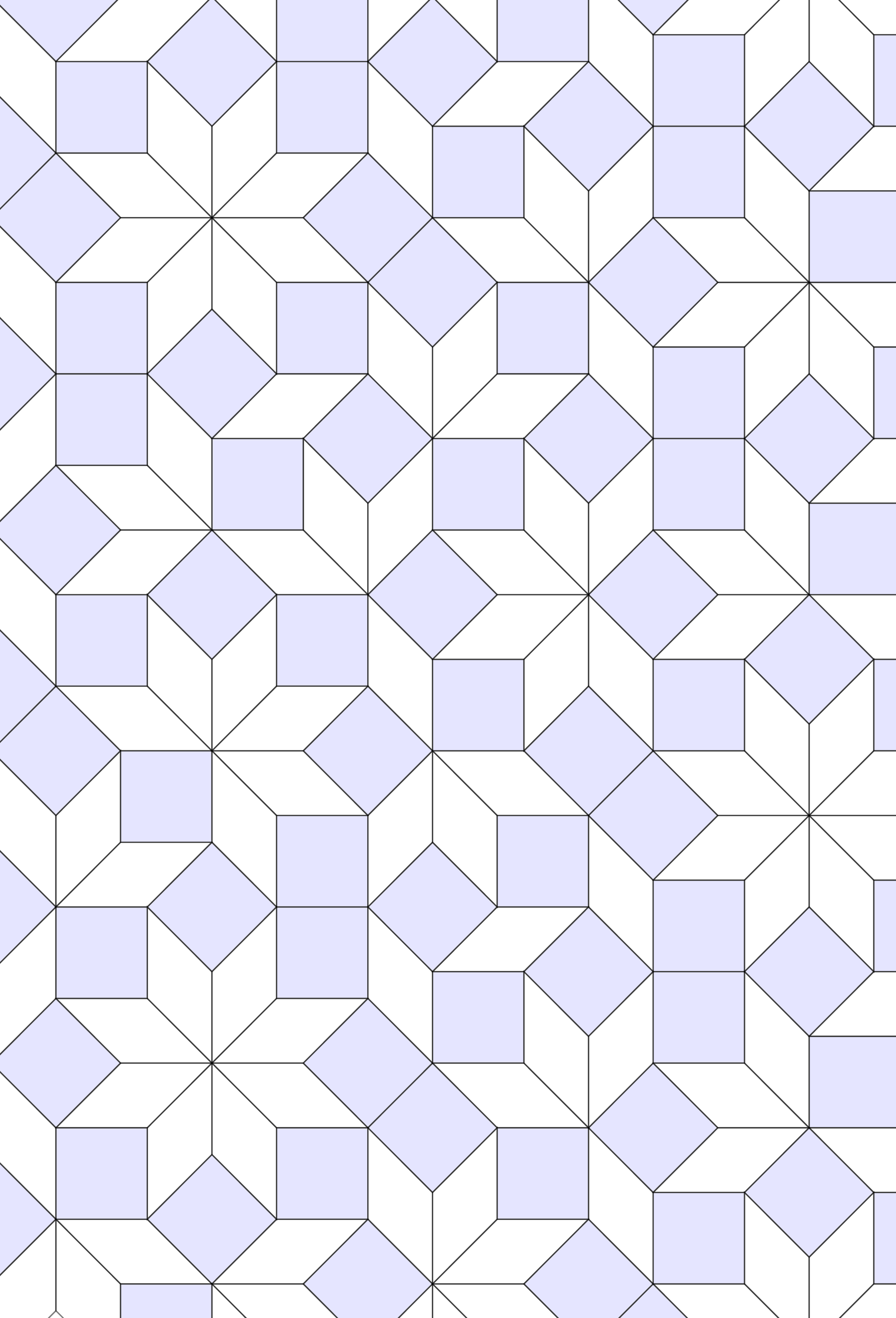}
    \hspace{2cm}
    \includegraphics[height=7cm]{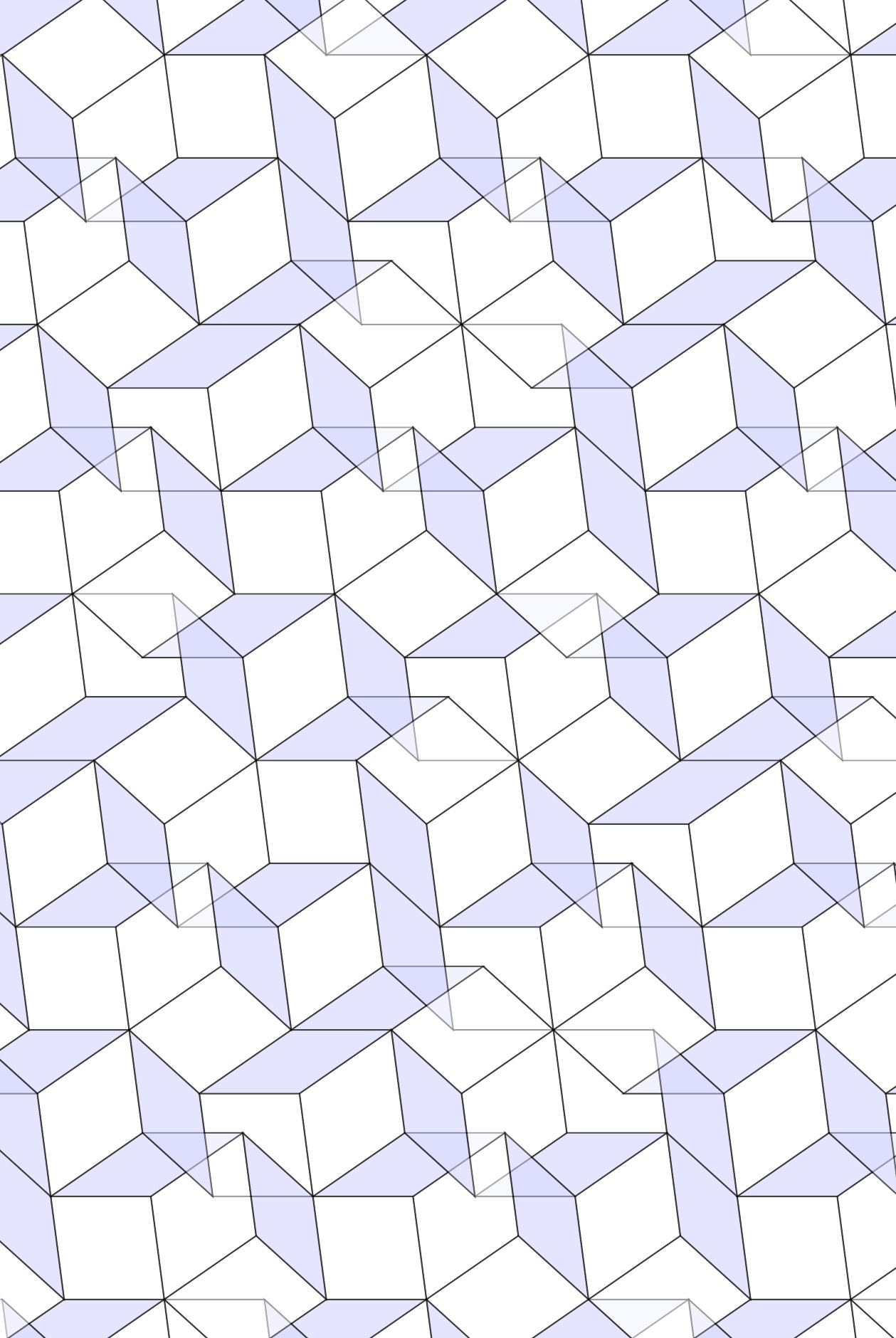}
    \caption{Golden octagonal tiling with the usual valid projection (left) and a non-valid projection on the same slope (right). Colors of the tiles are the same with respect to the $\pi(e_i)$'s, with an opacity of 50\% in both images.}
    \label{not-valid}
\end{figure}
Strongly planar tilings by parallelograms can also be obtained by the {\bf (canonical) cut and project method}.
For this, consider a $d$-dimensional affine plane $E\subset\R^n$  such that $E\cap\Z^n=\emptyset$, select (``cut'') all the $d$-dimensional facets of $\Z^n$ which lie within the tube $E+[0,1]^n$, then ``project'' them onto $\mathbb{R}^d$. If this projection $\pi$ yields a tiling of $\mathbb{R}^d$ it is called {\bf valid} (see Figure \ref{not-valid}), and the tiling is a strongly planar tiling by parallelograms with slope $E$.
Such tilings are called canonical cut and project tilings or simply \textbf{\textit{n}} $\to$ \textbf{\textit{d}} \textbf{tilings}. 
Not every projection is suitable, but the orthogonal projection onto $E$ seen as $\R^d$ is known to be valid \citep{harriss2004}. 
Here we only consider the case of a $2$-dimensional slope $E$ which is totally irrational, that is, which does not contain any rational line. 
This yields aperiodic tilings of the plane. \\

Figure \ref{ex1} illustrates the above notions with three well-known examples.
Rauzy tilings are $3\rightarrow 2$ tilings whose slope $E$ is generated by
$$
\vec{u}=(\alpha-1,-1,0)
\qquad\textrm{and}\qquad
\vec{v}=(\alpha^2-\alpha-1,0,-1),
$$
where $\alpha\approx 1.89$ is the only real root of $x^3-x^2-x-1$.
Ammann-Beenker tilings, composed of tiles of the set A5 in the terminology of \cite{grunbaum1987}, are the $4\rightarrow 2$ tilings with slope $E$ generated by
$$
\vec{u}=(\sqrt{2},1,0,-1)
\qquad\textrm{and}\qquad
\vec{v}=(0,1,\sqrt{2},1).
$$
Generalized Penrose tilings are the $5\rightarrow 2$ tilings with slope $E$ generated by
$$
\vec{u}=(\varphi,0,-\varphi,-1,1)
\qquad\textrm{and}\qquad
\vec{v}=(-1,1,\varphi,0,-\varphi),
$$
where $\varphi = (1 + \sqrt{5}) / 2$ 
is the golden ratio. 
The ``strict'' Penrose tilings as defined by Roger \cite{penrose1978} (set P3 in the terminology of \cite{grunbaum1987}) correspond to the case when $E$ contains a point whose coordinates sum to an integer.


\subsection{Local rules}
\label{sub:LR}

Local rules for tilings can be defined in several inequivalent ways.
Since we focus on cut and project tilings, we also define local rules for a slope.

Firstly, weak local rules for a tiling $T$ can be defined as in \cite{BF15}.
A \textbf{pattern} is a connected finite subset of tiles of $T$. 
Following \cite{Levitov1988}, an \textbf{r-map} of $T$ is a pattern formed by the tiles of $T$ which intersect a closed disk of radius $r\geq0$.
The \textbf{r-atlas} of $T$, denoted by $T(r)$, is then the set of all $r$-maps of $T$ (up to translation). In the case of a canonical cut and project tiling, it is a finite set.
A canonical cut and project tiling $\PP$ of slope $E$ is said to admit \textbf{weak local rules} if there exist $r\geq0$ and $t\geq1$, respectively called \textbf{radius} and \textbf{thickness}, such that any $n\to d$ tiling $T$ whose $r$-atlas is contained in $\PP(r)$ is planar with slope $E$ and thickness at most $t$.
By extension, the slope $E$ is then said to admit local rules.
In that case, we say that the slope of $\PP$ is characterized by its patterns of a given size.
Local rules are \textbf{strong} if $t=1$. 
Penrose tilings have strong local rules and the slope is characterized by patterns of the 1-atlas if the sides of the tiles have length 1 (see \cite{Senechal1995}, Theorem 6.1, p.177).

Another way of defining local rules is with Ammann bars. 
We call \textbf{Ammann segments} decorations on tiles which are segments whose endpoints lie on the borders of tiles, such that when tiling with those tiles, each segment has to be continued on adjacent tiles to form a straight line. 
We say that a slope $E$ admits \textbf{Ammann local rules} if there is a finite set of prototiles decorated with Ammann segments such that any tiling with those tiles is planar with slope $E$.
In particular, no periodic tiling of the plane should be possible with those tiles if $E$ is irrational.
For instance, the marking of the Penrose tiles yielding Ammann bars is shown in Figure \ref{segments}, along with a valid pattern where each segment is correctly prolonged on adjacent tiles.


\subsection{Subperiods}
\label{sub:subperiods}

\begin{figure}[ht!]
    \centering
    \begin{subfigure}{\textwidth}
    \includegraphics[width=0.95\textwidth]{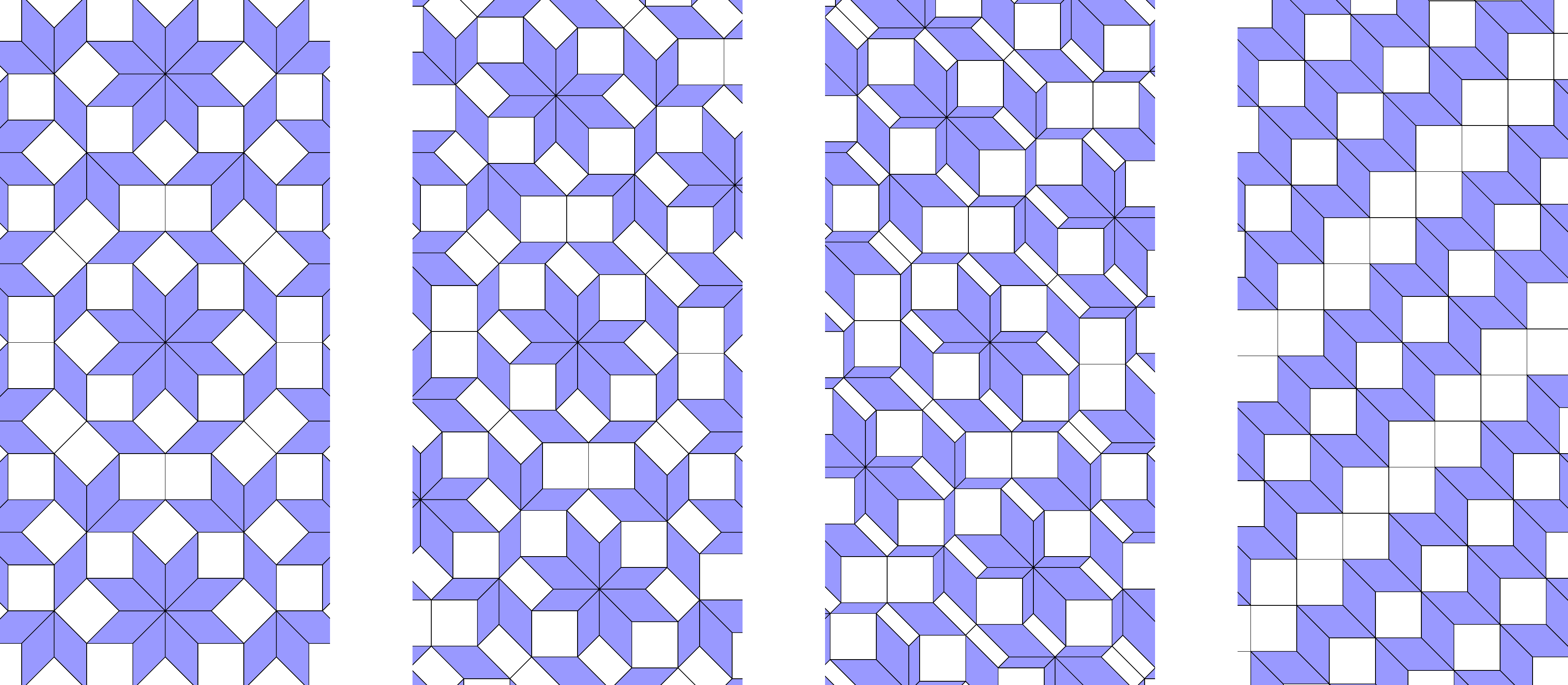}
    \caption{Starting from an Ammann-Beenker tiling (on the left), progressively reduce the length of one of the four vectors defining the tiles, until it is null (on the right). The shadow thus obtained is periodic in one direction.}
    \medskip
    \label{shadowAB}
    \end{subfigure}
    \begin{subfigure}{\textwidth}
    \includegraphics[width=0.95\textwidth]{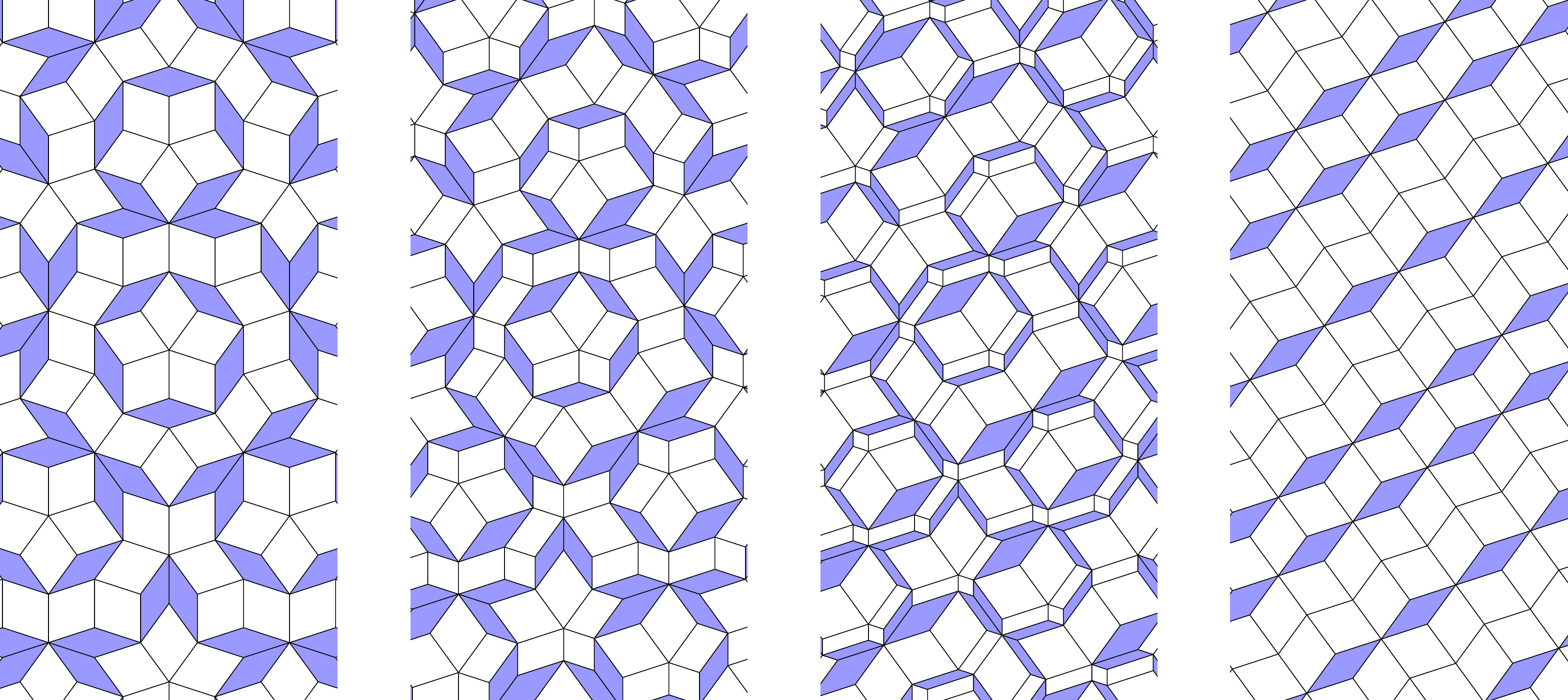}
    \caption{Starting from a Penrose tiling (on the left), progressively reduce the lengths of two of the five vectors defining the tiles, until they are null (on the right). The shadow thus obtained is periodic in one direction.}
    \label{shadowPenrose}
    \end{subfigure}
    \caption{Shadows of Ammann-Beenker and Penrose tilings.}
    \label{shadows}
\end{figure}

Adapted from \cite{BF2012}, the \textbf{$i_1,...,i_{n-3}$-shadow} of an $n\to 2$ tiling $T$ is the orthogonal projection $\omega_{i_1,...,i_{n-3}}$ of its lift on the space generated by $\{e_j\mid 0\leq j\leq n-1, j\neq i_1,...,i_{n-3}\}$.
This corresponds to reducing to zero the lengths of $\pi(e_{i_1}),...,\pi(e_{i_{n-3}})$ in the tiling, so that the tiles defined by these vectors disappear.
This is illustrated in Figure \ref{shadows}. 
An $n\to 2$ tiling thus has $n\choose 3$ shadows.

An \textbf{$i_1,...,i_{n-3}$-subperiod} of an $n\to 2$ tiling $T$ is a prime period of its $i_1,...,i_{n-3}$-shadow, hence an integer vector in $\R^3$.
By extension, we call subperiod of a slope $E$ any vector of $E$ which projects on a subperiod in a shadow of $T$.
A subperiod is thus a vector of $E$ with 3 integer coordinates: those in positions $j\notin\{i_1,...,i_{n-3}\}$.
We say that a slope is \textit{determined} or \textit{characterized} by its subperiods if only finitely many slopes have the same subperiods (in their shadows).
For instance, the slope of Ammann-Beenker tilings has four subperiods:
\begin{eqnarray*}
p_{0} &=& (\sqrt{2},1,0,-1),\\
p_{1} &=& (1,\sqrt{2},1,0),\\
p_{2} &=& (0,1,\sqrt{2},1),\\
p_{3} &=& (-1,0,1,\sqrt{2}).
\end{eqnarray*}
while that of Penrose tilings has ten, each with two non-integer coordinates.

This notion was first introduced by \cite{Levitov1988} as the \textit{second intersection condition} and then developed by \cite{BF15,BF2017}, who showed that in the case of $4\rightarrow2$ tilings, a plane admits weak local rules if and only if it is determined by its subperiods.
It was shown by \cite{BF2012} that this is not the case for Ammann-Beenker tilings: indeed, their subperiods are also subperiods of all Beenker tilings (introduced by \cite{Bee82}), that are the planar tilings with a slope generated, for any $s\in(0,\infty)$, by
$$
u=(1,2/s,1,0)
\qquad\textrm{and}\qquad
v=(0,1,s,1).
$$
The Ammann-Beenker tilings correspond to the case $s=\sqrt{2}$ and do not admit local rules.
On the other hand, generalized Penrose tilings have a slope characterized by its subperiods \citep{BF15} and do admit local rules.

In this article, we focus on $4\rightarrow2$ tilings with irrational slope $E$ characterized by four subperiods. In this case, each subperiod of $E$ has exactly one non-integer coordinate.
Since the vertices of the tiling are projected points of $\Z^4$, we define ``integer versions'' of subperiods: if $p_{i}=(x_0,x_1,x_2,x_3)$ is a subperiod, then its \textbf{floor} and \textbf{ceil} versions are respectively $\lfloor p_i\rfloor=(\lfloor x_0\rfloor,\lfloor x_1\rfloor,\lfloor x_2\rfloor,\lfloor x_3\rfloor)$ and $\lceil p_i\rceil=(\lceil x_0\rceil,\lceil x_1\rceil,\lceil x_2\rceil,\lceil x_3\rceil)$. 
Note that only the non-integer coordinate $x_i$ is affected, and that $\lfloor p_i\rfloor, \lceil p_i\rceil\notin E$.

\section{The FP-method}
\label{sec:FP-method}

In this section, we present a construction to get Ammann bars for some $4\rightarrow2$ tilings and we illustrate it with tilings that we call \textit{Cyrenaic tilings}.

\subsection{Fine projections}\label{sub:fineproj}

\begin{figure}
    \centering
    \includegraphics[width=0.6\textwidth]{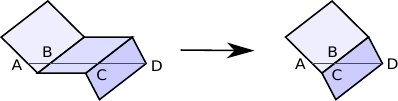}
    \caption{Aligned segments in a pattern remain aligned in the shadow corresponding to the direction of the line.}
    \label{continuity}
\end{figure}

In Subsection \ref{sub:canonical}, we defined the valid projections for a slope $E$ and mentioned the classical case of the orthogonal projection.
There are however other valid projections, and this will play a key role here.
Indeed, we will define Ammann bars as lines directed by subperiods and it will be convenient for the projected $i$-th subperiod $\pi(p_i)$ to be collinear with $\pi(e_i)$, 
so that the image of a line directed by $p_i$ is still a line in the $i$-th shadow (Figure \ref{continuity}).
This leads us to introduce the following definition:

\begin{definition}\label{def1}
A {\bf fine projection} for a $2$-dimensional slope $E\subset\R^4$ is a valid projection $\pi:\R^4\to\R^2$  such that for every $i\in\{0,1,2,3\}$, $\pi(p_i)$ and $\pi(e_i)$ are collinear.
\end{definition}

\begin{figure}[h!]
    \centering
    \includegraphics[height=6.7cm]{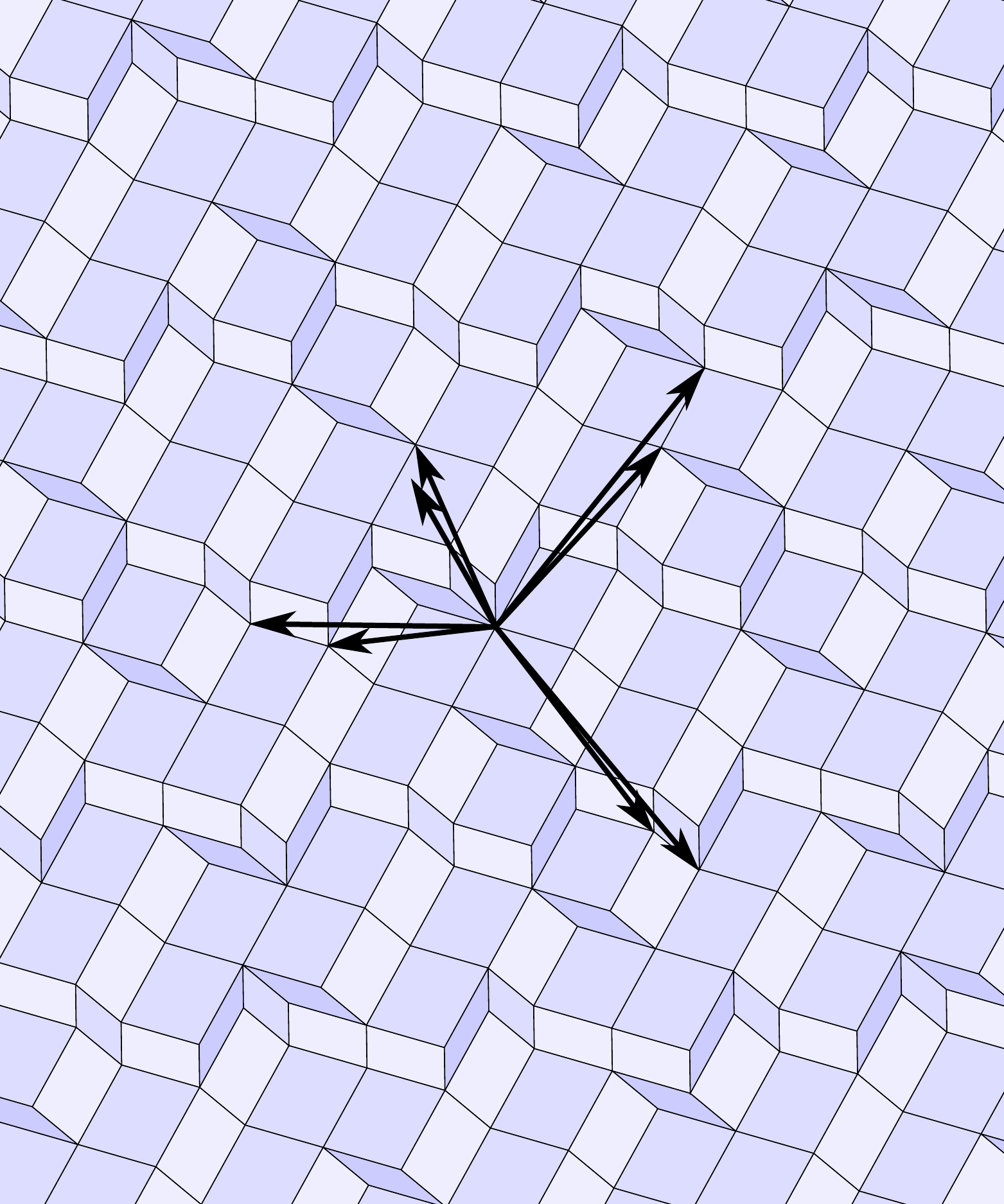}
    \qquad
    \includegraphics[height=6.7cm]{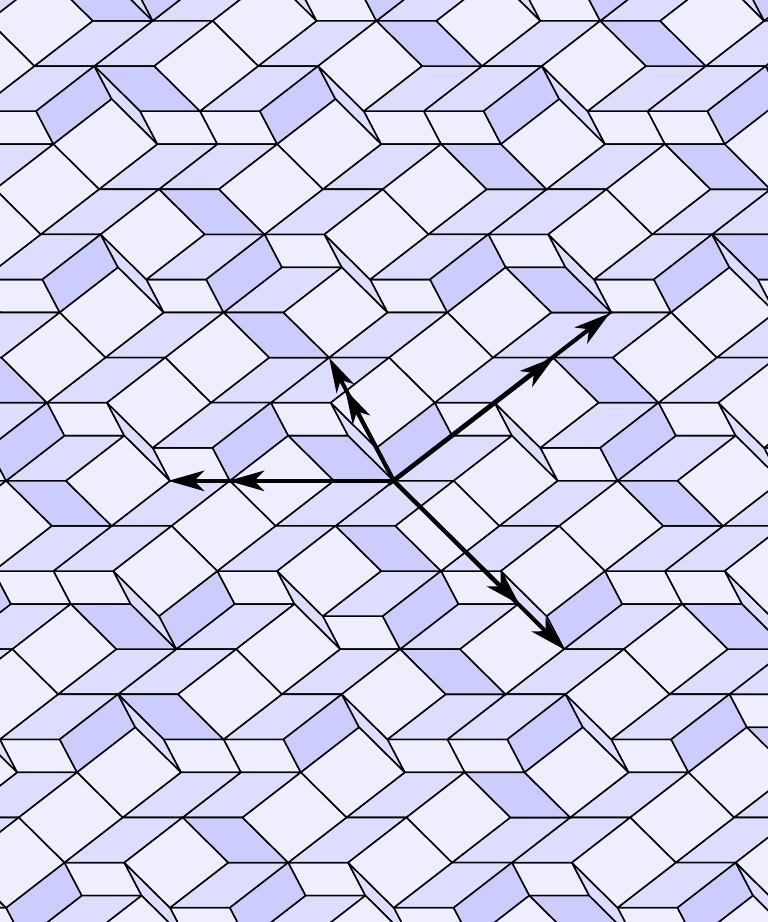}
    \caption{Cyrenaic tiling with $\pi(\lfloor p_i\rfloor)$ and $\pi(\lceil p_i\rceil)$ for each subperiod $p_i$. On the left, we used the orthogonal projection which is valid but not \textit{fine}; on the right we used a fine projection. Colors of the tiles are the same on both images with respect to the $\pi(e_i)$'s. Starting from the central pattern, one can see how one tiling is merely a deformation of the other (a rotation was applied). Note that from a given vertex (and a given $i$), sometimes only a translation through either $\pi(\lfloor p_i\rfloor)$ or $\pi(\lceil p_i\rceil)$ maps it to another vertex of the tiling, but not both.}
    \label{Cyrenaic}
\end{figure}

Figure \ref{Cyrenaic} illustrates the difference between two valid projections, one being fine but not the other, on the slope of Cyrenaic tilings which we present in the next subsection. With the fine projection, projected subperiods have the same directions as the sides of the tiles.
This is why, if segments on the tiles of a tiling $T$ are directed by $\pi(p_i)$, then continuity of the lines in direction $i$ is preserved in the $i$-shadow of $T$, for any $i\in\{0,1,2,3\}$, as illustrated in Figure \ref{continuity}.
Indeed, consider a line $L$ in direction $i$, then it is parallel to the sides of the tiles which disappear in the $i$-shadow of $T$.
Now consider a tile $t_0$ which disappears in this shadow, containing a segment $[BC]\subset L$, and its neighbors $t_{-1}$ and $t_1$ containing segments $[AB], [CD]\subset L$.
Taking the $i$-shadow corresponds to translating remaining tiles in direction $i$, hence by such a translation the endpoint of an Ammann segment is mapped to a point on the same line (namely the image of the other endpoint of the same segment).
As a result, the images of points $B$ and $C$ are on the same line, so that points $A, B, C, D$ are still aligned.

\subsection{Finding fine projections}
\label{sub:finding-fineproj}

Given a slope $E$ with subperiods $p_0,\ldots,p_3$, we search for a fine projection $\pi$ as follows.
We will define it by its $2\times4$ matrix $A$, which must
satisfy $Ae_i=\lambda_i Ap_i$ for $i=0,\ldots,3$, where $\Lambda:=(\lambda_i)_{i=0,\ldots,3}$ is to be determined.
With $M$ denoting the $4\times4$ matrix whose $i$-th column is $e_i-\lambda_ip_i$, this can be rewritten as $AM=0$.
The $2$ rows of $A$ must thus be in the left kernel of $M$.
Since the image of the facets in $E+[0,1]^4$ must cover $\R^2$, $A$ must have rank $2$.
Hence the left kernel of $M$ must be of dimension at least $2$, that is, $M$ must have rank at most $2$.
This is equivalent to saying that all the $3\times3$ minors of $M$ must be zero.
Each minor yields a polynomial equation in the $\lambda_i$'s.
Any solution of the system formed by these equations yields a matrix $M$ whose left kernel can be computed.
If the kernel is not empty, then any basis of it yields a suitable matrix $A$.

Of course with 4 variables and 16 equations there is no guarantee that a solution exists, 
and oftentimes when a projection respects the collinear condition in Definition \ref{def1} it is not valid: some tiles are superimposed in what should be a tiling. 
Figure \ref{not-valid} shows for instance what happens in the case of golden octagonal tilings (introduced in \cite{BF15}) when the obtained matrix $A$ is used.
To find a slope $E$ with a fine projection, we proceed as follows:
\begin{enumerate}
    \item Randomly choose the three integer coordinates of each subperiod $p_i$;
    \item Check that only finitely many slopes admit these subperiods;
    \item Use the above procedure to find a fine projection (if any);
    \item Repeat until a fine projection is found.
\end{enumerate}

To get an idea of the proportion of irrational slopes for which a fine projection exists, we ran some statistics with the following procedure.
For an integer parameter $k$, we randomly choose a quadratic polynomial $P$ with coefficients in $[-k,k]$ until we find an irreducible one, then two irrational numbers $x$ and $y$ in the quadratic field defined by $P$.
We also randomly choose two vectors $u$ and $v$ in $\Z^4$ with coefficients in $[-k,k]$, and we replace the first coordinate in $u$ by $x$ and the second in $v$ by $y$.
Finally, we apply the above method on the slope $E$ generated by $u$ and $v$.
For $k=3$, out of 999 slopes, 222 were not characterized by subperiods and 116 admitted a fine projection.
For $k=5$, out of 7015 slopes, 765 were not characterized by subperiods and 1000 admitted a fine projection\footnote{We searched until we found 1000 fine projections.}.\\

Among the first examples that we found, the following caught our attention because it has very short subperiods.
Here are the integer coordinates of these:
\begin{eqnarray*}
p_0 &=& (*,0,1,1),\\
p_1 &=& (1,*,-1,1),\\
p_2 &=& (1,-1,*,0),\\
p_3 &=& (2,1,-1,*),
\end{eqnarray*}
where $*$ stands for the non-integer coordinate.
We checked that there are only two ways to choose these non-integer coordinates so that the subperiods indeed define a plane, with $a=\pm\sqrt{3}$:
\begin{eqnarray*}
p_0 &=& (a,0,1,1),\\
p_1 &=& (1,a-1,-1,1),\\
p_2 &=& (1,-1,a+1,0),\\
p_3 &=& (2,1,-1,a).
\end{eqnarray*}
Proceeding as explained at the beginning of this subsection yields
$$
M=\frac{1}{6}\left(\begin{array}{ccccccc}
3 & & -a & & -a & & -2 a \\
0 & & a + 3 & & a & & -a \\
-a & & a & & -a + 3 & & a \\
-a & & -a & & 0 & & 3
\end{array}\right),
$$
whose left kernel is generated, for example, by the rows of the matrix
$$
A:=\frac{1}{2}\left(\begin{array}{ccccccc}
2 & & 0 & & a + 1 & & a - 1 \\
0 & & 2 & & -a - 1 & & a + 1
\end{array}\right)
$$
Only $a=\sqrt{3}$ defines a valid projection, so we choose this value.
We denote by $E_c$ the slope generated by the $p_i$'s and call {\bf Cyrenaic tilings} the $4\to 2$ tilings with slope $E_c$.
Figure \ref{Cyrenaic} illustrates this.

\subsection{Defining the prototiles}

We describe here the FP-method, which we used to obtain for instance the tileset $\CC$ depicted in Figure~\ref{decotiles}.
\begin{figure}[hbt]
    \centering
    \includegraphics[width=0.49\textwidth]{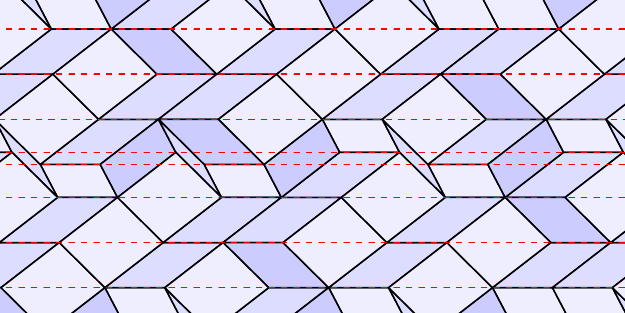}
    \hfill
    \includegraphics[width=0.49\textwidth]{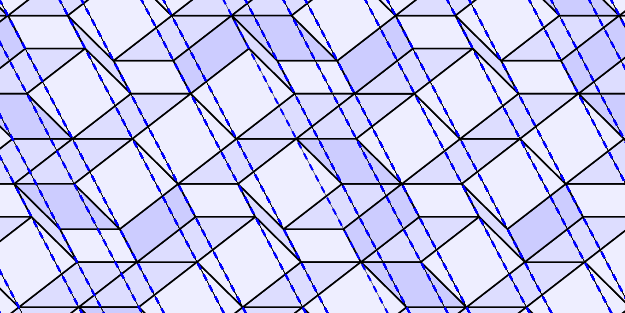}\\
    \smallskip
    \includegraphics[width=0.49\textwidth]{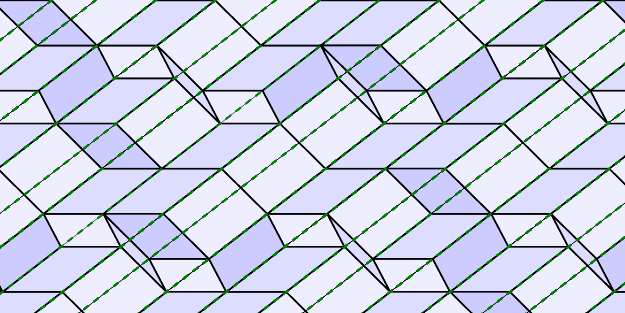}
    \hfill
    \includegraphics[width=0.49\textwidth]{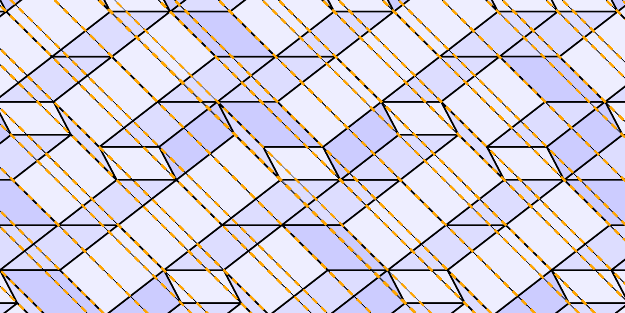}
    \caption{A Cyrenaic tiling with all the lines in the directions of the subperiods, through every vertex of the tiling. Directions are shown separately to ease visualization, and lines are dashed so that one can see the edges of the tiling.}
    \label{lines}
\end{figure}
Let $E$ be a $2$-dimensional irrational plane in $\R^4$ characterized by its subperiods and which admits a fine projection $\pi$.
Consider a tiling with slope $E$ obtained using the fine projection $\pi$.
Draw through each vertex of this tiling four lines directed by each of the projected subperiods $\pi(p_i)$'s.
Figure~\ref{lines} shows what we obtain for a Cyrenaic tiling.
These lines decorate the tiles of the tiling with segments that can take four different directions.
All these decorated tiles, considered up to translation, define the wanted tileset.
Note that the tileset does not depend on the initially considered tiling, because the $4\to 2$ tilings with a given irrational slope share the same finite patterns (this known fact is e.g. proven by Prop. 1 in \cite{BF2017}).
We can now prove:

\begin{proposition}\label{prop:finite_tileset}
Any tileset obtained using the FP-method is finite.
\end{proposition}

\begin{figure}[h!]
    \centering
    \includegraphics[width=0.5\textwidth]{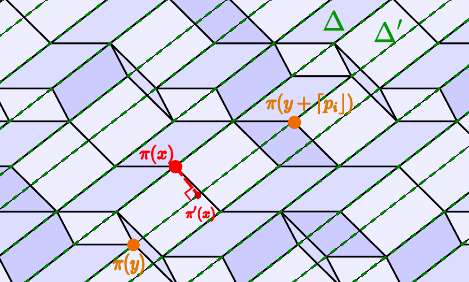}
    \caption{Illustration of the Proof of Proposition \ref{prop:finite_tileset}. $\lceil p_i\rfloor$ stands for $\lfloor p_i\rfloor$ or $\lceil p_i\rceil$.}
    \label{dessin-prop1}
\end{figure}

\begin{proof}
We prove that the number of different intervals (distances) between two consecutive lines in a given direction is finite.
This yields finitely many ways to decorate a tile by parallel segments, hence finitely many different tiles.
Consider a subperiod $p_i$ and the set $\mathcal{D}_i$ of all lines in $E$ directed by $\pi(p_i)$ and passing through the vertices of the tiling, that is by all points $\pi(x)$ with $x\in\Z^4\cap(E+[0,1]^4)$.
Since the distance from a vertex to its neighbors is $||\pi(e_k)||$ for some $k$, the interval between two consecutive lines of $\mathcal{D}_i$ is at most $d_1:=\max_{j\neq i}\{||\pi(e_j)||\}$.

Let $\Delta\in\mathcal{D}_i$, $x\in\R^4$ such that $\pi(x)\in\Delta$, and $\Delta'\in\mathcal{D}_i$ which is closest to $\Delta$ (Fig. \ref{dessin-prop1}). 
Then the distance from $\pi(x)$ to its orthogonal projection $\pi'(x)$ on $\Delta'$ is at most $d_1$.
Next, the distance between two vertices lying on $\Delta'$ is at most $d_2:=\max(||\pi(\lfloor p_i\rfloor)||,||\pi(\lceil p_i\rceil)||)$. 
Indeed, if $y\in\Z^4\cap(E+[0,1]^4)$ then $y+p_i\in E+[0,1]^4$ and has three integer coordinates so that it lies on an edge of $\Z^4$ (seen as a grid in $\R^4$), between $y+\lfloor p_i\rfloor$ and $y+\lceil p_i\rceil$; now at least one of these two points is in $\Z^4\cap(E+[0,1]^4)$, therefore its projection is also a vertex of the tiling, which lies on $\Delta'$ (since $\pi(p_i)$, $\pi(\lfloor p_i\rfloor)$ and $\pi(\lceil p_i\rceil)$ are collinear).
Hence the distance between $\pi'(x)$ and the closest vertex $\pi(y)$ of the tiling which lies on $\Delta'$ is at most $d_2/2$.
As a result, $\dist(\pi(x),\pi(y))\leq d:=\sqrt{d_1^2+d_2^2/4}$, i.e at least one vertex on $\Delta'$ is in the ball $B(\pi(x),d)$.
Consequently, measuring the intervals around a line $\Delta$ in the $d$-maps of the tiling is enough to list all possible intervals between two consecutive lines in the whole tiling. 
Since the $d$-atlas is finite, so is the number of intervals.
\hfill $\square$%
\end{proof}
\\

Although the previous proof does not give an explicit bound on the number of tiles, it does give a constructive procedure to obtain these tiles.
It is indeed sufficient to compute the constant $d$ (which depends on the subperiods and the projection), then to enumerate the $d$-maps (for example by enumerating all patterns of size $d$ and keeping only those which can be lifted in a tube $E+[0,1]^4$ -- in practice we used a more efficient algorithm based on the notion of region \citep{BF2020} described in Appendix \ref{annexe-code} -- and, for each $d$-map, to draw the lines and enumerate the new decorated tiles obtained.
In the case of Cyrenaic tilings, it is sufficient to enumerate the tiles which appear in the 5-atlas in terms of graph distance
.
We obtain 2 or 3 intervals in each direction, and the set $\CC$ of 36 decorated prototiles in Figure \ref{decotiles}.

\section{Tiling with an FP-tileset}\label{sec:tileset}

By construction, an FP-tileset obtained from a slope $E$ characterized by subperiods can be used to form all the canonical cut and project tilings of slope $E$.
However, even respecting the Ammann bars rules, nothing yet ensures that these tiles cannot be used to tile in other ways, and obtain for instance tilings which would be periodic or not planar.
We shall here prove that this actually cannot happen.

Say we have a set $S$ of tiles decorated with Ammann segments obtained from a given slope $E\subset\R^4$ characterized by subperiods $(p_i)_{i\in\{0,1,2,3\}}$ with a fine projection $\pi$, and we want to show that any tiling with those tiles is planar with slope $E$.
Let $\T$ be the set of all tilings that can be made with (only) tiles of $S$.
By construction (assembly rules for the tiles in $S$), four sets of lines appear on any $T\in\T$ and the lines of each set are parallel to a projected subperiod $\pi(p_i)$ and to $\pi(e_i)$ for the same $i$.
We can therefore talk about the $i$-shadow of $T$ as the lift in $\R^3$ of the tiling obtained when reducing to zero the length of sides of tiles which are parallel to $\pi(p_i)$.
Then as shown in Subsection \ref{sub:fineproj}, for any $i\in\{0,1,2,3\}$, continuity of the lines in direction $i$ is preserved in the $i$-shadow of $T$.

We can then use the lines to show that a shadow is periodic (in one direction) and determine its prime period: starting from a vertex of the shadow, we follow the line in the chosen direction until we hit another vertex, for each valid configuration of the tiles. 
If the vector from the first vertex to the next is always the same, then it is a prime period of the shadow.

\begin{lemma}\label{lemma}
  Suppose that $E$ is a 2-dimensional slope in $\R^4$ characterized by subperiods $p_0,...,p_3$, $\pi$ is a fine projection, and $\T$ is a tileset obtained from $E$ and $\pi$ using the FP-method. Then for any tiling composed with tiles of $\T$ (respecting Ammann bars rules), its $i$-shadow is $\omega_i(p_i)$-periodic for all $i\in\{0,1,2,3\}$.
\end{lemma}
\begin{proof}
For more clarity, the diagram in Figure \ref{diagram} summarizes the notations  and the ``traveling'' between spaces that we use.
\begin{figure}[h!]
    \centering
    \begin{tikzpicture}
        \draw (0.5,0) node {Tiling ($\R^2$)} ;
        \draw (0.5,2) node {Lift in $\R^4$} ;
        \draw (6,2) node {$i$-shadow in $\R^3$} ;
        \draw (6,0) node {Projected $i$-shadow ($\R^2$)} ;
        \draw[->] (2,0) -- (4,0);
        \draw (3,0) node[below] {$\Omega_i$} ;
        \draw[->] (0.5,1.5) -- (0.5,0.5);
        \draw (0.5,1) node[left] {$\pi$} ;
        \draw[->] (2,2) -- (4,2);
        \draw (3,2) node[above] {$\omega_i$} ;
        \draw[->] (6.2,1.5) -- (6.2,0.5);
        \draw (6.2,1) node[right] {$\pi_i$} ;
        \draw[->] (5.8,0.5) -- (5.8,1.5);
        \draw (5.8,1) node[left] {$\wedge$} ;
    \end{tikzpicture}
    \caption{Diagram summarizing notations used in Lemma \ref{lemma}.}
    \label{diagram}
\end{figure}
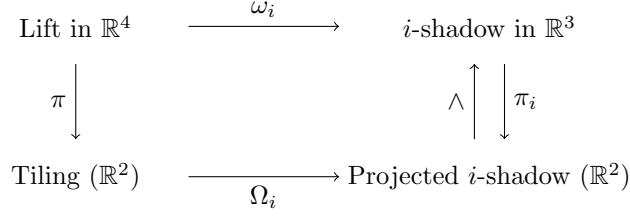
Starting from the lift (in $\R^4$) of a tiling of $\R^2$, projecting through $\pi$ brings us back to the tiling.
As in Subsection \ref{sub:subperiods}, $\omega_i$ denotes the projection of the lift along $e_i$.
This corresponds to shrinking to zero the length of each edge of type $\pi(e_i)$ in the tiling, which is done by applying $\Omega_i$ to the tiling.
This yields the projected $i$-shadow, which is also a tiling of $\R^2$ and can be obtained from the $i$-shadow through the projection $\pi_i$.
When the tileset is constructed using our algorithm, we could add a bit more information on the tiles: each endpoint of a decoration (intersection of a line with a side of a tile) can be identified by two points $\hat{x}$ and $\Bar{x}$ in $\R^3$ as described in steps 1 and 2 of this proof.
Note that this ``additional'' information is entirely determined by a point $x$ which is already encoded in the prototile.
Hence it is only useful to show the result in step 3 of this proof, but we do not need to actually encode it.\\
\textbf{Step 1. }
For steps 1 and 2, we consider the $i$-shadow of a ``good'' tiling, namely the canonical cut-and-project tiling used to obtain the decorated prototiles, with the lines added on it.
So we know that its $i$-shadow $S_i$ is periodic.
Starting from $\pi_i(S_i)$, which is a tiling of $\R^2$, its lift in $\R^3$ is $S_i$ and the lift of a point $y$ in this tiling is $\hat{y}\in S_i$ such that $\pi_i(\hat{y})=y$, i.e. $\hat{y}$ is on a 2-dimensional facet of $\R^3$ whose vertices (integer points) are in the shadow and project on vertices of a tile $t$ in the projected $i$-shadow .
Now consider a point $y$ in a copy of $t$, and its position $x$ in $t$.
Then $\hat{x}:=\hat{y}\bmod\Z^3$ is the position of $\hat{y}$ in the facet of the shadow which projects on $t$, and it does not depend on the position of the tile in the tiling.
Furthermore, we have $\pi_i(\hat{x})=x$.
\begin{figure}[h]
    \centering
    \begin{subfigure}[ht]{0.48\textwidth}
    \includegraphics[width=\textwidth]{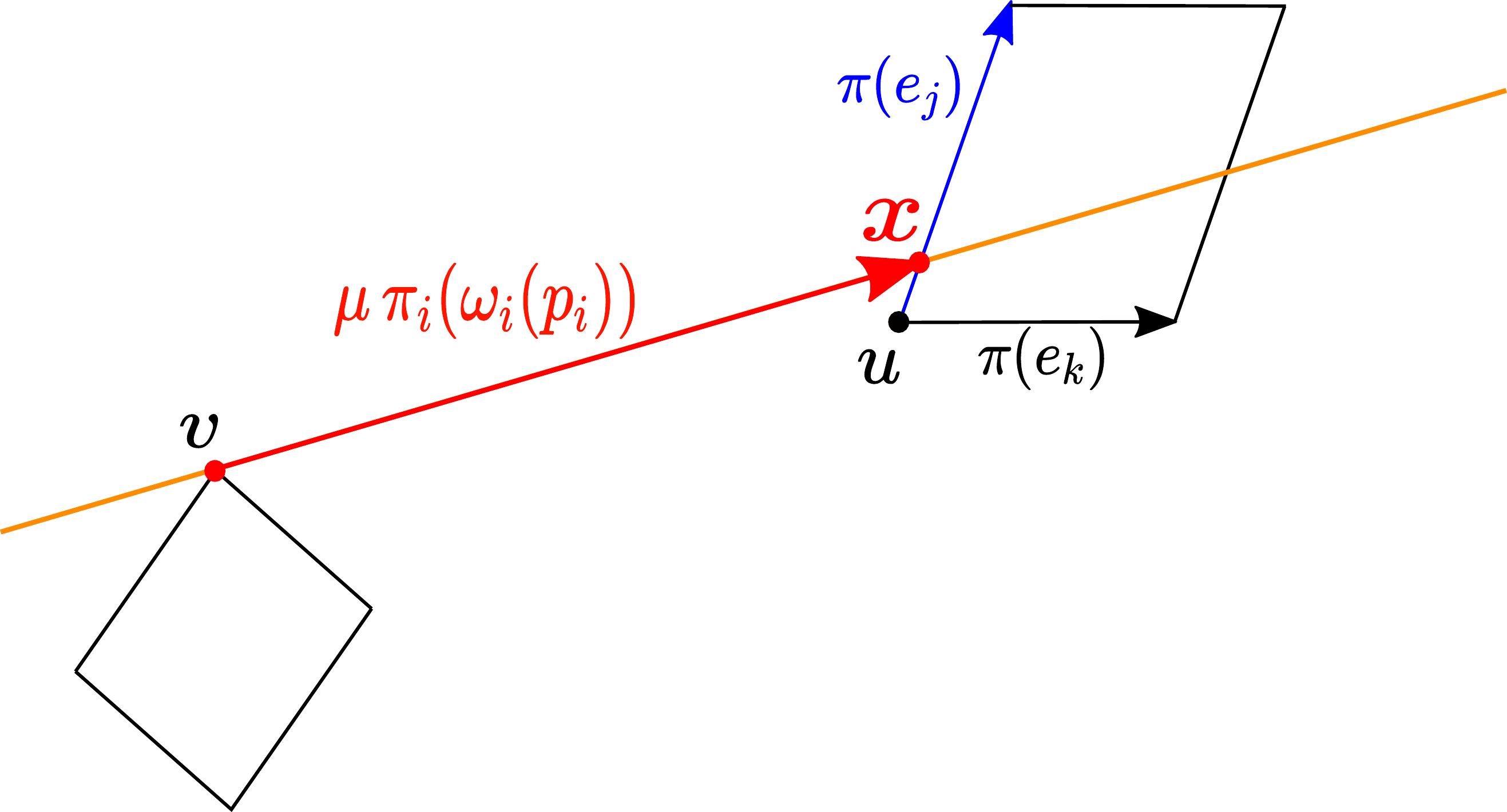}
    \caption{In the tiling ($\R^2$)}
    \end{subfigure}
    \begin{subfigure}[ht]{0.48\textwidth}
    \includegraphics[width=\textwidth]{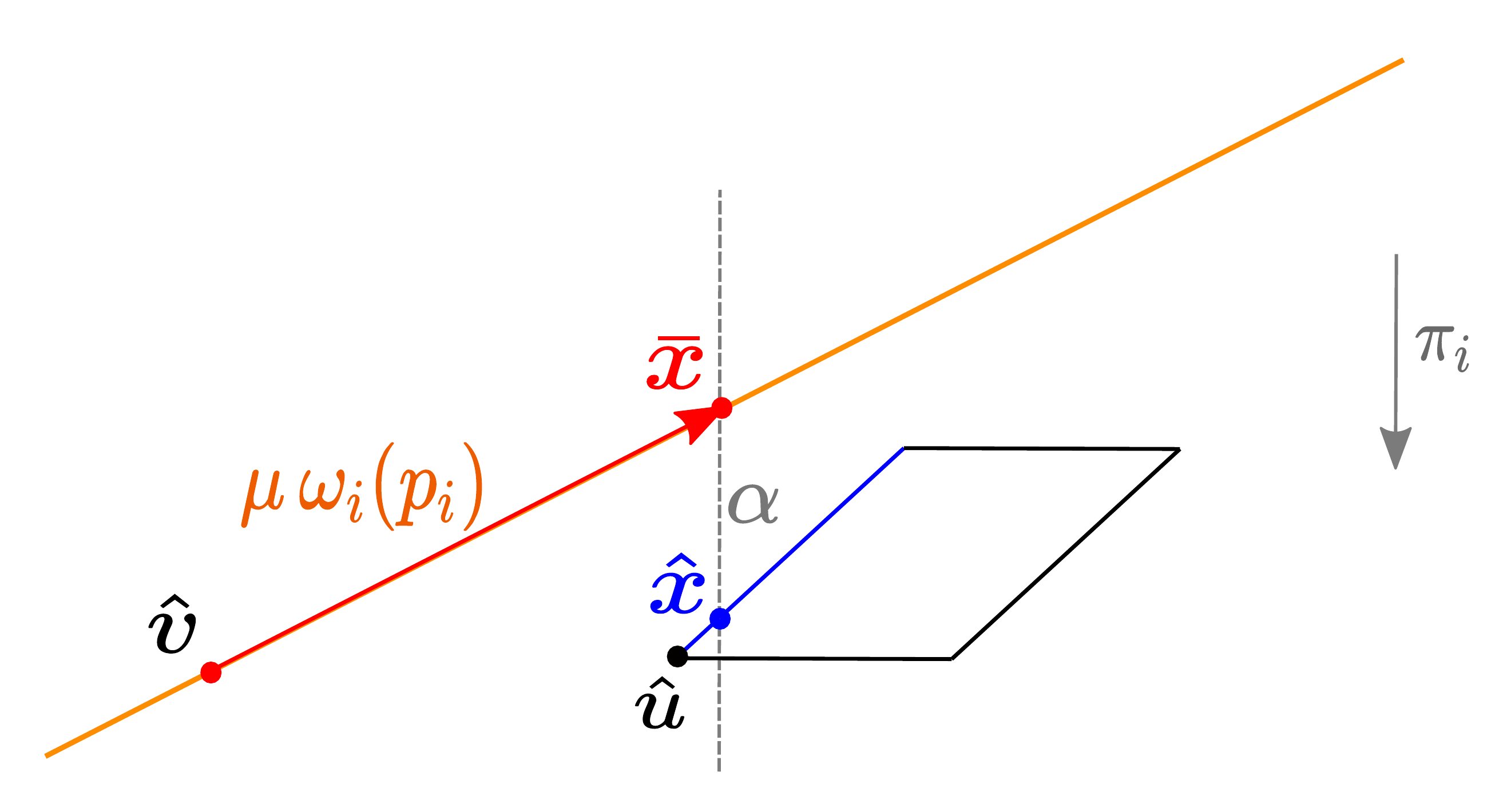}
    \caption{In the $i$-shadow ($\R^3$)}
    \end{subfigure}
\end{figure}
\\
\textbf{Step 2. } 
Besides, we lift the lines in $\R^3$ as follows.
If $y$ is on a line directed by $\pi_i(\omega_i(p_i))$, by construction there is a vertex $v$ on this line. 
Consider the line$\Delta$ in $\R^3$ directed by $\omega_i(p_i)$ through $\hat{v}$. 
The projection of this line is the line where $y$ lies, so there exists $\Bar{y}\in\Delta$ such that $\pi_i(\Bar{y})=y$. Hence we have $\Bar{y}=\hat{v}+\mu\,\omega_i(p_i)$ with $\mu\in\R$.
Then $x$ is mapped to $\Bar{x}:=\mu\,\omega_i(p_i)\mod\Z^3=\Bar{y}\bmod\Z^3$.
We show that $\Bar{x}$ is entirely determined by $x$, i.e. does not depend on the position of the tile in the tiling.
Indeed, $\ker(\pi_i)$ has dimension 1, thus $\ker(\pi_i)=\R q_i$ with $q_i\in\R^3$ and $||q_i||=1$.
Since $\pi(\hat{x})=\pi(\Bar{x})=x$, we have $\hat{x}-\Bar{x}=\alpha q_i$ with $\alpha\in\R$, therefore $\hat{x}-\Bar{x}\equiv \hat{x}-\mu\,\omega_i(p_i)\equiv \alpha q_i\bmod\Z^3$,
which is equivalent to $\alpha q_i+\mu\,\omega_i(p_i)\equiv \hat{x}$, which does not depend on $x$.
Moreover $\pi_i$ is a valid projection, hence $q_i\wedge\omega_i(p_i)\neq0$, so $\alpha$ and $\mu$ are determined by $x$.
In other words, no matter which copy of each tile yields the prototile in the tileset, for a given position $x$ in the prototile $\Bar{x}$ will be the same.
We also observe that $\Bar{x}=0$ if and only if $x$ is a vertex of the tile. Indeed, if $\Bar{x}=0$ then $\mu\in\Z$ because $\omega_i(p_i)$ is an integer vector whose coordinates are relatively prime. And by definition $\Bar{x}=\hat{v}+\mu\,\omega_i(p_i)\bmod\Z^3$ where $v$ is a vertex of the tiling, so $\hat{v}$ is a vertex of the shadow and it follows that the $i$-shadow is $\omega_i(p_i)$-periodic, thus $\Bar{y}$ is also a vertex of the shadow, hence $y$ is a vertex of the tiling.\\
\textbf{Step 3. }
We finally show that the $i$-shadow of \textit{any} tiling composed with the tiles in $\T$ is periodic with the right period.
We now consider a tiling composed only using the tiles of $\T$ which can appear in the projected $i$-shadow, respecting Ammann bars rules for the lines directed by $\pi(p_i)$ (continuity of the lines cannot be respected in other directions with only those tiles).
If $y$ is a vertex of this tiling, then for the corresponding $x$ we have $\Bar{x}=0$. 
Following the line from $y$ in direction $i$, we reach a point $z$ in the tiling.
Then since $\omega_i(p_i)\in\Z^3$, we have $\Bar{z}=\Bar{x+\pi_i(\omega_i(p_i))}=\Bar{x}+\omega_i(p_i)\equiv0\bmod\Z^3$.
As a result, $z$ is also a vertex of the shadow, which is thus $\omega_i(p_i)$-periodic.
\hfill $\square$
\end{proof}

For instance, every tiling composed with tiles of $\CC$ has the same subperiods as Cyrenaic tilings.
If the above proof seems opaque, here is an example of what happens in the shadows.
We observe that each $i$-shadow is periodic with period $q_i:=\omega_i(p_i)$ where $p_i$ is the $i$-subperiod of Cyrenaic tilings.
This is shown in Figure \ref{CC-shadows}. 
\setcounter{figure}{10}
\begin{figure}[ht]
    \centering
    \begin{subfigure}[ht]{0.52\textwidth}
    \includegraphics[width=\textwidth]{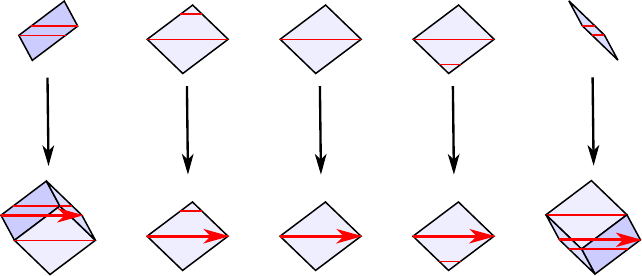}
    \caption{0-shadow}
    \end{subfigure}
    \hspace{1.5cm}
    \vspace{0.3cm}
    \begin{subfigure}[ht]{0.28\textwidth}
    \includegraphics[width=\textwidth]{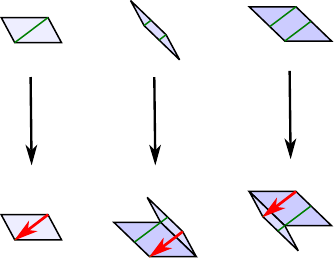}
    \caption{2-shadow}
    \end{subfigure}
    \vspace{0.3cm}
    \begin{subfigure}[ht]{0.8\textwidth}
    \includegraphics[width=\textwidth]{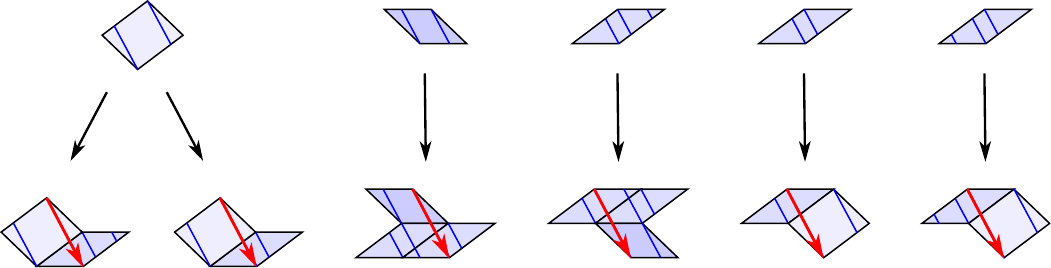}
    \caption{1-shadow}
    \end{subfigure}
    \begin{subfigure}[ht]{\textwidth}
    \includegraphics[width=\textwidth]{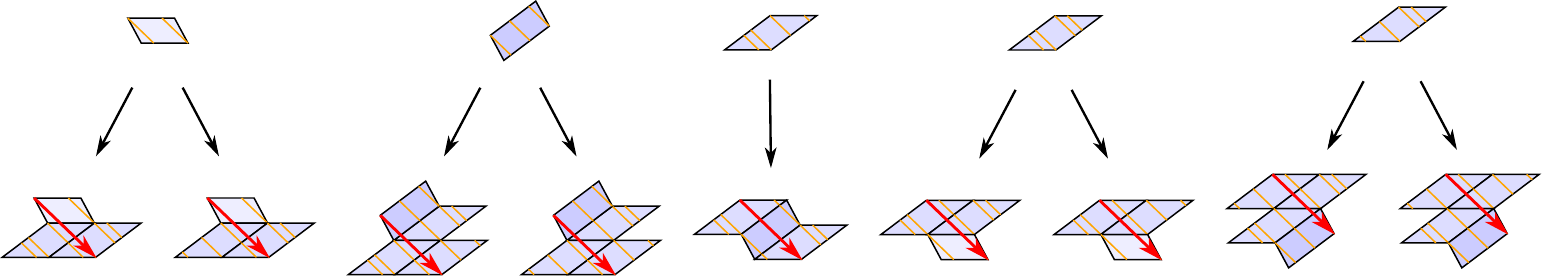}
    \caption{3-shadow}
    \end{subfigure}
    \caption{Periods of the 4 shadows of tilings that can be realized with the set $\CC$: starting at any vertex and following a line in direction $i$, depending on the first traversed tile, there are at most two possibilities until reaching another vertex, and the vector between both vertices is always the same.}
    \label{CC-shadows}
\end{figure}
In each shadow there are three original (non-decorated) tiles, each of which can appear in different versions when taking the decorations into account.
For each $i$-shadow here we only look at the decorations in direction $i$, where we have the continuity of the lines (other decorations are irrelevant here).
All possible tiles are given on the top row, and following the arrows from each tile one can see all of the different options\footnote{Remember that a line passes through every vertex, in each direction.} for placing other tiles in order to continue the line in the direction of the red vector.
For each shadow, the vector is the same for all possible configurations, which means that the shadow is periodic, and we find exactly the subperiods of Cyrenaic tilings.\\

The main result in \cite{BF15} thus yields the following:

\begin{corollary}
Every tiling composed with tiles of $\T$ is planar with slope $E$.
\end{corollary}

For a totally irrational slope $E$, Theorem \ref{thm} follows.

Note that there is no guarantee their thickness is always $1$.
Indeed, it is difficult in general to understand the thickness of a lifted tiling.
Moreover, our result relies on that of Bédaride and Fernique which does not guarantee a thickness of 1.
Even in the Cyrenaic case, we are not sure whether Ammann decorations enforce strong planarity: we have a result on shadows and the forcing seems strong, but it is possible sometimes to ``flip'' between $\pi(\lfloor p_i\rfloor)$ and $\pi(\lceil p_i\rceil)$ to jump from one vertex to another, and this might yield a thickness $>1$ in some cases, so we would have to examine what can happen or not in the tiling.
Therefore, we prefer not to claim anything at this point.

\bigskip
\noindent {\bf Acknowledgements. }
We would like to thank Alexandre Blondin Massé and the five reviewers of LATIN 2022 for their careful proofreading of the conference version of this article, and the reviewer of this long version for their very interesting remarks and relevant suggestions to improve the presentation of our work.

\appendix
\section{Ammann bars and subperiods in Penrose tilings}
\label{annexe-penrose}
We present here in more detail the links between Ammann bars and subperiods in Penrose tilings.

\subsection{Subperiods}
\label{sub:penrose-sub}

The definition of subperiods is given in Subsection \ref{sub:subperiods}, here are those of Penrose tilings, which have the form $p_{ij}=(x_0,x_1,x_2,x_3,x_4)$ with $x_i,x_j\in\R$ and $x_k\in\Z$ for $k\notin\{i,j\}$:
\begin{center}
\begin{tabular}{lp{1.5cm}l}
$p_{01} = (1-\varphi,\varphi-1,1,0,-1)$,&&
$p_{02} = (-\varphi,0,\varphi,1,-1)$,\\
$p_{03} = (\varphi,1,-1,-\varphi,0)$,&&
$p_{04} = (\varphi-1,1,0,-1,1-\varphi)$,\\
$p_{12} = (1,\varphi-1,1-\varphi,-1,0)$, &&
$p_{13} = (1,\varphi,0,-\varphi,-1)$,\\
$p_{14} = (0,\varphi,1,-1,-\varphi)$,&&
$p_{23} = (0,1,\varphi-1,1-\varphi,-1)$,\\
$p_{24} = (1,-1,-\varphi,0,\varphi)$,&&
$p_{34} = (1,0,-1,1-\varphi,\varphi-1)$.\\
\end{tabular}
\end{center}
For the following, all indices are in $\F_5$ (all operations are modulo 5) and we also define some ``integer versions'' of subperiods, with $i,j\in\{0,1,2,3,4\}, i<j$:
\begin{itemize}
    \item ``ceil-floor'' $\lceil p_{ij}\rfloor$ is $p_{ij}$ with $\lceil x_i\rceil$ in place of $x_i$ and $\lfloor x_j\rfloor$ in place of $x_j$,
    \item ``floor-ceil'' $\lfloor p_{ij}\rceil$ is $p_{ij}$ with $\lfloor x_i\rfloor$ in place of $x_i$ and $\lceil x_j\rceil$ in place of $x_j$.
\end{itemize}

Firstly, we observe that for each subperiod, $\pi(\lceil p_{ij}\rfloor)$ and $\pi(\lfloor p_{ij}\rceil)$ have the same direction as $\pi(p_{ij})$ ($\pi$ denotes the orthogonal projection onto the slope).
More precisely, 
\begin{itemize}
    \item the shortest of $\pi(\lfloor p_{i,i+1}\rceil)$ and $\pi(\lceil p_{i,i+1}\rfloor)$ (the first one except for $i=3$) is equal to $\pi(e_{\text{min}(i-1,i+2)})-\pi(e_{\text{max}(i-1,i+2)})$ which has the same direction as $\pi(e_i)-\pi(e_{i+1})$ and corresponds to the long diagonal of a thin rhombus, and the other is $\varphi$ times longer;
    \item the shortest of $\pi(\lfloor p_{i,i+2}\rceil)$ and $\pi(\lceil p_{i,i+2}\rfloor)$ is equal to $\pm\varphi(\pi(e_{i})-\pi(e_{i+2}))$ which corresponds to the short diagonal of a thick rhombus, and the other is $\varphi$ times longer (which corresponds to the long diagonal of a thick rhombus).
\end{itemize}
For instance, $\lfloor p_{14}\rceil = (0,1,1,-1,-1)$, $\lceil p_{14}\rfloor = (0,2,1,-1,-2)$, and for $a\in\R$ we have
$$(0,1+a,1,-1,-(1+a)) = a e_1+(e_1+e_2)-(e_3+e_4) -a e_4$$
Hence, since $\pi(e_i+e_{i+1})=-\varphi \pi(e_{i-2})$ for all $i\in\F_5$,
\begin{align*}
\pi(0,1+a,1,-1,-(1+a)) 
    &= a\pi(e_1)-\varphi\pi(e_4) + \varphi\pi(e_1)-a\pi(e_4)\\
    &= (\varphi+a)(\pi(e_1)-\pi(e_4))
\end{align*}
Thus for $a=0,\varphi-1,1$ we obtain that $\pi(\lfloor p_{14}\rceil)$, $\pi(p_{14})$ and $\pi(\lceil p_{14}\rfloor)$ are collinear. Only the order of coordinates changes for other subperiods.

\subsection{Link with Ammann bars}


Firstly, directions of the subperiods are the same as directions of the Ammann bars, $\pi(p_{ij})$ being orthogonal to $\pi(e_i)+\pi(e_j)$. For instance, $p_{01}\bot e_3$ and we have $\pi(e_0)+\pi(e_1)=-\varphi \pi(e_3)$ since the long diagonal of a thick rhombus is $\varphi$ times longer than the side
. 
Figure \ref{segments} illustrates these relations, which are the same in every direction up to rotation: it shows part of a Penrose tiling with Ammann bars as well as $\lfloor p_{ij}\rceil$ and $\lceil p_{ij}\rfloor$ for one subperiod. 
Moreover, when the side of a tile has length $\varphi$, on the one hand the ``distances'' between Ammann bars of a given set, measured not orthogonally but with an angle of $\frac{2\pi}{5}$ rad = 72 degrees
, are $S=2\sin\frac{2\pi}{5}$ and $L=\varphi S$; and on the other hand the norms of $\pi(\lceil p_{i,i+1}\rfloor)$ and $\pi(\lfloor p_{i,i+1}\rceil)$ are $L$ and $S+L=\varphi L$ i.e. $\varphi$ times longer (the long diagonal of a thin rhombus has length $L$).

Note that there is exactly one Ammann segment in every direction in each tile. 
Hence taking into account the distances between consecutive bars, respecting conditions on Ammann bars forces subperiods.

\section{Details about computations}
\label{annexe-code}

Reading this appendix is not necessary to understand the article but may be of interest to readers who find that too many calculations are swept under the rug.
Some of these calculations are difficult and rely on functions that we wrote using SageMath (Python with classical mathematical functions that are very useful here). All these functions can be found at the following address :
\begin{center}
    \url{https://github.com/cporrier/Cyrenaic}
\end{center}
The purpose of this appendix is to explain how these functions work.

\subsection{Computing cut and project tilings}

We use the duality between {\em multigrids} \cite{bruijn1981} and cut and project tilings.
A {\em grid} is a set of regularly spaced parallel hyperplanes (lines in the case of an $n\to 2$ tiling).
A multigrid is then a $d\times n$ matrix $G$ whose rows define the direction and spacing of each grid (normal vector to each hyperplane, with the norm giving the spacing), and a {\em shift vector} $S$ which specifies how each grid is translated away from the origin.
The shift must be chosen such that no more than $d$ hyperplanes intersect in a point (this is generic).

The function \texttt{generators\_to\_grid(E)} convert a slope $E$, given by $d$ vectors of $\R^n$ which generate it, to a multigrid $G$.

Then, to each intersection of $d$ hyperplanes corresponds a tile: it is generated by the directions of the hyperplanes, and the $i$-th coordinate of its position is the number of hyperplane in the $i$-th direction between the origin and the considered intersection.

The function \texttt{dual(G,S,k)} computes the {\em dual} of the multigrid $G$ with shift vector $S$ and $2k+1$ grids in each direction (since we can only compute a subset of the infinite tiling).
This is a set of tiles represented each by a pair $(t,pos)$ where $t$ is the $d$-tuple of indices of the vectors of the standard basis which define the prototile and $pos$ is the integer translation applied on it (in $\Z^n$).

\subsection{Finding the integer entries of a subperiod}

Consider a slope $E$ generated by the vectors $u_1,...,u_d\in \R^n$.
We assume that the entries of the $u_i$'s are in $\Q(a)$ for some algebraic number $a$, because it has been proven in \cite{BF2020} to be a necessary condition to have local rules.
We want to find a subperiod $p$ of $E$, that is a vector of $E$ with $d+1$ integer coordinates -- the \textit{type} of $p$ gives us the indices of its integer coordinates $a_{i_1},...,a_{i_{d+1}}$ ($1\leq i_1<i_2<...<i_{d+1}\leq n$).
For this, we consider the $n\times(d+1)$ matrix $N$ whose columns are the $u_j$'s and the subperiod $p$ to be determined:
$$N=\left(\begin{array}{c|c|c|c|c}
 &  &  &  & {\color{red}.} \\
{\color{red}.} & {\color{red}.} & {\color{red}\ldots} & {\color{red}.} & {\color{red}a_{i_1}} \\
{\color{red}.} & {\color{red}.} & {\color{red}\ldots} & {\color{red}.} & {\color{red}a_{i_2}} \\
 &  &  &  & {\color{red}\vdots} \\
u_1~ & ~u_2~ & ~\cdots~ & ~u_d~ & p \\
{\color{red}.} & {\color{red}.} & {\color{red}\ldots} & {\color{red}.} & {\color{red}a_{i_k}} \\
 &  &  &  & {\color{red}\vdots} \\
{\color{red}.} & {\color{red}.} & {\color{red}\ldots} & {\color{red}.} & ~{\color{red}a_{i_{d+1}}} \\
 &  &  &  & {\color{red}.} 
\end{array}\right)$$

Since $p\in E$, this matrix has rank $d$ and all the $(d+1)$-minors are thus zero.
In particular, the nullity of the minor obtained by selecting the lines $i_1,\ldots,i_d$ yields a linear equation on the $a_{i_k}$.
Indeed, developing the minor along the last column yields
$$
(a_{i_1},...,a_{i_{d+1}})\times
\left(\begin{array}{c}
(-1)^1G_{i_2...i_{d+1}}\\
 \vdots \\
(-1)^kG_{i_1...\widehat{i_k}...i_{d+1}}\\
 \vdots \\
(-1)^{d+1}G_{i_1...i_d}\\
\end{array}\right)=0,
$$
where $G_{i_1\ldots\widehat{i_k}\ldots i_{d+1}}$ is the determinant of the $d\to d$ matrix obtained by taking the coordinates $i_1,\ldots i_{d+1}$ except $i_k$ of the $u_i$'s (it is also known as {\em Grassmann coordinates} of the plane).
Since the $a_i$'s must be integer, this can equivalently be rewritten as follows.
Replace each coefficient $(-1)^kG_{i_1...\widehat{i_k}...i_{d+1}}$ by a line of length $\textrm{deg}(a)$, the algebraic degree of $a$, whose $i$-th entry is the coefficient of $a^i$ in $(-1)^kG_{i_1\ldots\widehat{i_k}\ldots i_{d+1}}$.
This yields a $n\times \textrm{deg}(a)$ integer matrix.
Then, the $a_i$'s are obtained by computing the left kernel of this matrix.

The above algorithm is implemented in the function \texttt{subperiods(E)}, which outputs a list of subperiods represented each by its type and its $d+1$ integer entries.

Let us illustrate this with the Cyrenaic tiling.
The slope $E$ is generated by the lines of the matrix
$$
\left(\begin{array}{cccc}
a & 0 & 1 & 1 \\
1 & a - 1 & -1 & 1
\end{array}\right),
$$
where $a=\sqrt{3}$.
Let us search a subperiod whose first three entries are integer.
The matrix $N$ is
$$
N=
\left(\begin{array}{ccc}
a & 1 & a_1\\
0 & a-1 & a_2\\
1 & -1 & a_3\\
1 & 1 & *
\end{array}\right),
$$
where $*$ denotes the non-integer entry we are here not interested in.
Consider the $3$-minor obtained with the three first lines and develop it along the last column.
We get:
$$
(a_1,a_2,a_3)\times
\left(\begin{array}{r}
-a + 1 \\
a + 1 \\
-a + 3
\end{array}\right)=0.
$$
Since $a=\sqrt{3}$, $\textrm{deg}(a)=2$ and this is equivalent to
$$
(a_1,a_2,a_3)\times
\left(\begin{array}{rr}
-1 & 1 \\
1 & 1 \\
-1 & 3
\end{array}\right)=0.
$$
This  $3\times 2$ integer matrix turns out to have rank $2$.
Its left kernel has dimension $1$ and the $a_i$'s are given by a prime integer vector in this kernel:
$$
(a_1,a_2,a_3)=(2, 1, -1).
$$
This yields the subperiod $p_3=(2,1,-1,*)$, as claimed in Subsection~\ref{sub:finding-fineproj}.

\subsection{Determining whether a slope is characterized by its subperiods}

To determine whether a slope $E$ is characterized by its subperiods, we first compute the subperiods of $E$ as explained in the previous subsection.
Then, we form the matrix whose columns are these subperiods, with variables $x_i$'s for the non-integer coordinates (since only the integer coordinates are known).
Since these subperiods must belong to the $d$-dimensional plane $E$, any $(d+1)$-minor of this matrix must be zero.
This yields an equation.
By considering all the minors, we get a system of polynomial equation.
The slope is characterized by its subperiods if and only if this system has dimension zero.

This is implemented in the function \texttt{is\_determined\_by\_subperiods(E)}.

\subsection{Finding the non-integer entries of a subperiod}

To compute the subperiods of a slope $E$ seen as vectors in $E$, we first compute the integer entries of the subperiods as explained above. We then consider the unknown non-integer entries has variables $x_i$'s. The matrix formed by the subperiods has rank $d$ since all the subperiods must be in the $d$-dim. plane $E$. Its $(d+1)$-minors must thus all be zero. This yields a polynomial system of equations in the $x_i$'s. We take the solution which is in $E$: this yields all the coordinates of the subperiods.

This is implemented in the function \texttt{lifted\_subperiods(E)}.

\subsection{Finding a fine projection}

The method to find a fine projection is described in Subsection \ref{sub:finding-fineproj}.
It is implemented by the function \texttt{fine\_projection(E)}.
We also implemented the function \texttt{valid\_projection(A,E)}, which checks whether the projection $A$ is valid for the slope $E$.

\subsection{Computing the \textit{r}-atlas}

Computing the $r$-atlas of a cut and project tiling mainly relies on the notions of {\em window} and {\em region}.
We here briefly recall these notions; the interested reader can find more details in \cite{BF2020}.

Consider a strongly planar $n\to d$ tiling with a given slope $E$.
Consider the simplest pattern: a single edge directed by $\pi(e_i)$.
To decide whether this pattern appears somewhere in the tiling, we have to decide whether there exists a vertex $x$ of the tiling such that $x+\pi(e_i)$ in also a vertex of the tiling.
By definition, a vertex $x$ belongs to the tiling if and only if its lift $\widehat{x}$ belongs to the tube $E+[0,1]^n$.
The idea is to look in the space orthogonal to $E$, denoted by $E'$ (sometimes called {\em internal space}, while $E$ is called {\em real space}).
Denote by $\pi'$ the orthogonal projection onto $E'$.
Now, a vertex $x$ belongs to the tiling iff $\pi'(\widehat{x})$ belongs to the polytope $W:=\pi'([0,1]^n)$.
This polytope is called the {\em window} of the tiling.
Similarly, $x+\pi(e_i)$ belongs to the tiling iff $\pi'(\widehat{x}+e_i)$ belongs to $W$, that is, iff $\pi'(\widehat{x})$ belongs to $W-\pi'(e_i)$.
Hence, there exists two vertices $x$ and $x+\pi(e_i)$ of the tiling if and only if the following polytope is not empty:
$$
R(e_i):=W\cap (W-\pi'e_i).
$$
The polytope $R(e_i)$ is called the {\em region} of the pattern formed by a single edge directed by $\pi(e_i)$.
This can be extended to any pattern $P$: such a pattern appears somewhere in the tiling iff its region, defined as follows, is not empty:
$$
R(P):=\bigcap_{x\in P}W-\pi'(\widehat{x}),
$$
where the intersection is taken over the vertices of $P$.
This is easily implemented in the function\break \texttt{region(W,ip,P)}, which takes as parameter the window, the internal projection and the pattern.\\

Let us now explain how to use regions to compute the $r$-atlas.
We shall maintain a list of the already computed $r$-maps, together with their regions in the window $W$.
We start with an empty list and fill it progressively as follows.

While the already computed regions do not cover the whole $W$, we first pick at random a point $z$ in $W$ which is not in one of the already computed region.
We then associate with $z$ the set of points $u\in \Z^n$ with norm at most $r$ such that $z+\pi'u\in W$.
This set is an $r$-map, which is new because its region does not overlap the already computed regions.
We compute the region of this $r$-map and we add both the map and the region in our list.

There is an additional tricky minor detail.
Before computing the region of an $r$-map, we must ``close'' this $r$-map, that is, add the tiles that are forced by the $r$-map: whenever two consecutive segments on the boundary of the pattern form a notch where a single tile can be added, then we must add this tile (because we know that there is no other segments dividing this notch - the information that there is nothing is indeed an information on its own).

All this is implemented in the function \texttt{atlas(E,r)}, which use two auxiliary functions.

\subsection{Computing the decorated tiles}

To compute the decorated tiles, we first compute a sufficiently large atlas (in the case of the Cyrenaic tilings the $6$-atlas is sufficient).
Then, for each pattern in this atlas, we compute the lines directed by the subperiods which go through each vertex of the pattern and intersect them with the tile at the origin of the pattern: this yields one decorated tile.
This is implemented in the function \texttt{decorated\_tile}.
We proceed similarly for each pattern of the atlas (function \texttt{decorated\_tiles}) to get the whole decorated tileset.

\bibliography{tilings}

\end{document}